\documentclass[useAMS,usenatbib,usegraphicx]{mn2e}

\usepackage{epsfig}
\usepackage{amsmath,amssymb}
\usepackage{natbib}
\usepackage{graphicx}
\usepackage{siunitx}
\newcommand{\Te} {$T_{\rm eff}$}
\newcommand{\logg} {$\log g$}
\newcommand{\loghe} {$\log$~(He/H)}
\newcommand{\msun} {M$_{\odot}$}
\newcommand{\rsun} {R$_{\odot}$}
\newcommand{\halpha} {H$\alpha$}
\newcommand{\kms} {km~s$^{-1}$}

\title[Ultracool WDs]{Ultracool white dwarfs and the age of the Galactic disc\thanks{This work is based on observations
obtained at the MDM Observatory, operated by Dartmouth College,
Columbia University, Ohio State University, Ohio University,
and the University of Michigan.}}
\author[A. Gianninas et al.]
{\parbox{\textwidth}{A. Gianninas$^{1}$\thanks{E-mail: alexg@nhn.ou.edu}, 
B. Curd$^{1}$,  
John R. Thorstensen$^{2}$, Mukremin Kilic$^{1}$, \\
P. Bergeron$^{3}$, Jeff J. Andrews$^{4}$, Paul Canton$^{1}$, and M. A. Ag\"ueros$^{4}$}\vspace{0.4cm}\\
\parbox{\textwidth}{
$^{1}$Homer L. Dodge Department of Physics \& Astronomy, University of Oklahoma, 440 W. Brooks St, Norman, OK 73019, USA\\
$^{2}$Department of Physics and Astronomy, Dartmouth College, 6127 Wilder Laboratory, Hanover, NH 03755, USA\\
$^{3}$D\'epartement de Physique, Universit\'e de Montr\'eal, C.P. 6128, Succ. Centre-Ville, Montr\'eal, QC H3C 3J7, Canada \\
$^{4}$Department of Astronomy, Columbia University, 550 West 120th Street, New York, NY 10027, USA \\
}}

\begin{document}
\date{Accepted 2015 March 8. Received 2015 March6 1; in original form 2015 February 4}

\pagerange{\pageref{firstpage}--\pageref{lastpage}} \pubyear{2015}

\maketitle

\label{firstpage}

\begin{abstract}
We present parallax observations and a detailed model atmosphere
analysis of 54 cool and ultracool (\Te\ $<$~4000~K) white dwarfs (WDs)
in the solar neighbourhood. For the first time, a large number of cool and
ultracool WDs have distance and tangential velocities measurements
available. Our targets have distances ranging from 21~pc to
$>$~100~pc, and include five stars within 30~pc. Contrary to
expectations, all but two of them have tangential velocities smaller
than 150~\kms\ thus suggesting Galactic disc membership. The
oldest WDs in this sample have WD cooling ages of 10~Gyr, providing a
firm lower limit to the age of the thick disc population. Many of our
targets have uncharacteristically large radii, indicating that they
are low-mass WDs. It appears that we have detected the brighter
population of cool and ultracool WDs near the Sun. The fainter
population of ultracool CO-core WDs remain to be discovered in large
numbers. The Large Synoptic Survey Telescope
should find these elusive, more massive ultracool WDs in the solar
neighbourhood.
\end{abstract}

\begin{keywords}
techniques: photometric -- stars: atmospheres -- stars: evolution -- white dwarfs -- Galaxy: disc
\end{keywords}

\section{Introduction}
Given the finite age of the Universe, the first asymptotic giant
branch stars that formed now live as
$\log$~($L/{\rm L}_{\odot}$)~=~$-$4.5 white dwarfs
\citep[WDs;][]{mestel52,iben84,winget87,liebert88,fontaine01}.  Such WDs have
temperatures below 4000~K (hence classified as ultracool) and they
have been observed in deep {\em Hubble Space Telescope} ($HST$) images
of the halo globular clusters M4 and NGC~6397
\citep{hansen04,hansen07}.  The oldest WDs in these two clusters are
$\approx$11.5~Gyr old.

Large-scale surveys such as the Sloan Digital Sky Survey
\citep[SDSS;][]{gates04,harris06,harris08,kilic06a,kilic10a,vidrih07,hall08},
the UKIRT Infrared Deep Sky Survey
\citep[UKIDSS;][]{leggett11,catalan12,tremblay14} and SuperCOSMOS
\citep{hambly99,rowell08} have identified the analogues of these
ultracool WDs in the field. Since these field WDs are relatively
bright compared to the globular cluster WDs, optical and infrared
photometry in several bands can be easily obtained from ground-based
telescopes, enabling us to model their spectral energy distributions
(SEDs) accurately. This is important for understanding the different
opacity sources in these stars, deriving reliable temperatures and
ages, and also calibrating the faint WD sequences of globular clusters
that usually rely on two filter photometry.

The spectra of hydrogen-rich cool and ultracool WDs differ from those
of their warmer counterparts because they show the effects of the
red-wing of the Ly$\alpha$ opacity in the blue \citep{kowalski06} and
the collision-induced absorption (CIA) due to molecular hydrogen in
the near-infrared \citep{hansen99}. The latter shifts the peak of the
SEDs of ultracool WDs back to the optical wavelengths. Unfortunately,
there are only three ultracool WDs in the field with parallax
measurements. These are WD~0346+246\footnote{We note that this object
  is also known as WD~0343+247.}, SDSS~J110217.48+411315.4
\citep[hereafter J1102;][and references therein]{kilic12} and
LHS~3250 \citep{bergeron02}. The first two stars have SEDs that peak
near 1~$\mu$m. On the other hand, the LHS~3250 SED peaks at
0.6~$\mu$m, representing an extreme case of CIA flux deficit in the
optical and infrared. \citet{bergeron02} performed a detailed model
atmosphere analysis of LHS~3250 and demonstrated that LHS~3250 has a
helium-rich composition, it is overluminous, and undermassive. The
best-fitting model and the parallax measurement indicate a mass of only
0.23~\msun\ \citep{bergeron02}. This is somewhat problematic as all
previously known low-mass WDs are DAs with hydrogen-rich atmospheres.

\citet{gates04} and \citet{harris08} as well as several other groups
have identified about a dozen stars with SEDs similar to LHS~3250. In
this paper, we present parallax measurements and a model atmosphere
analysis of 54 cool WDs, including half a dozen ultracool WDs and
several other cool WDs with significant infrared flux deficits. Our
targets were selected from the cool and ultracool WD samples of
\citet{gates04}, \citet{vidrih07}, \citet{harris08} and
\citet{kilic10a}, and are biased towards WDs with significant infrared
flux deficits.  Parallax measurements allow us to accurately determine
the distances, masses and consequently the cooling ages for these
stars. Section 2 outlines our observations including a description of
our Bayesian approach to estimating distances. Section 3 describes the
models used in our analyses followed by our results in Section 4. In
Section 5, we discuss the ages and membership of the WDs in our sample
as well as the implications of our results towards our understanding
of WD evolution and we conclude in Section 6.

\section{Observations}

\subsection{Parallax}
All our parallax data are from the 2.4m Hiltner telescope at Michigan-Dartmouth-MIT (MDM)
Observatory on Kitt Peak, Arizona.  We used a thinned SITe CCD (named
`echelle'); at the $f 7.5$ focus, each $24\ \mu$m pixel subtended
0.275 arcsec, giving a field of view 9.4 arcmin$^{2}$.  For all our
parallax data, we used a 4-inch-square Kron--Cousins $I$-band filter,
which did not vignette the CCD.  Exposure times varied with the
brightness of the object, but were typically a few hundred seconds.
Our data were taken on numerous observing runs between 2007 and 2011.
Table~\ref{tab:obs} gives the epochs that each star was observed, and
the number of exposures at each epoch.

\begin{table*}
\centering
\begin{minipage}{0.975\textwidth}
\caption{Journal of parallax observations.}
\begin{tabular}{@{}lrrrl@{}}
\hline
\noalign{\smallskip}
SDSS     & $N_{\rm ref}$ & $N_{\rm meas}$ & $N_{\rm pix}$ & \multicolumn{1}{c}{Epochs} \\
\noalign{\smallskip}
\hline
\noalign{\smallskip}
J0045+1420    &   30  &  57  &  50  & 2007.73(4), 2007.82(3), 2008.69(11), 2008.88(16), 2008.97(8), 2009.72(8) \\
J0121$-$0038  &   15  &  48  & 115  & 2007.73(8), 2008.05(10), 2008.69(16), 2008.88(17), 2008.97(9), 2009.73(13), 2009.86(18), \\
              &       &      &      & 2010.01(10), 2011.75(9), 2011.94(5) \\
J0146+1404    &   35  &  54  & 107  & 2007.73(8), 2008.05(8), 2008.69(12), 2008.88(18), 2008.97(8), 2009.73(12), 2009.86(16), \\
              &       &      &      & 2010.01(12), 2011.75(3), 2011.94(10) \\
J0256$-$0700  &   15  &  41  & 149  & 2007.74(33), 2007.81(12), 2008.05(8), 2008.69(13), 2008.88(14), 2009.03(10), 2009.73(8), \\
              &       &      &      & 2009.86(12), 2010.02(7), 2011.75(10), 2011.93(22) \\
J0301$-$0044  &   25  &  58  & 102  & 2007.73(7), 2007.82(6), 2008.06(8), 2008.69(10), 2008.88(17), 2008.97(8), 2009.73(12), \\
              &       &      &      & 2009.86(13), 2010.01(12), 2011.75(9) \\
J0309+0025    &   17  &  47  & 126  & 2007.74(8), 2007.81(10), 2008.05(8), 2008.69(1), 2008.88(14), 2008.97(7), 2009.72(8), \\
              &       &      &      & 2009.86(16), 2010.02(13), 2011.75(16), 2011.93(25) \\
\noalign{\smallskip}
\hline
\noalign{\smallskip}
\multicolumn{5}{@{}p{\textwidth}@{}}{($Note.$ This table is available in its entirety 
in a machine-readable form in the online journal. A portion is shown here for 
guidance regarding its form and content.)} \\
\label{tab:obs}
\end{tabular}
\end{minipage}
\end{table*}

\begin{figure}
\centering
\includegraphics[scale=0.575,bb=90 32 542 769]{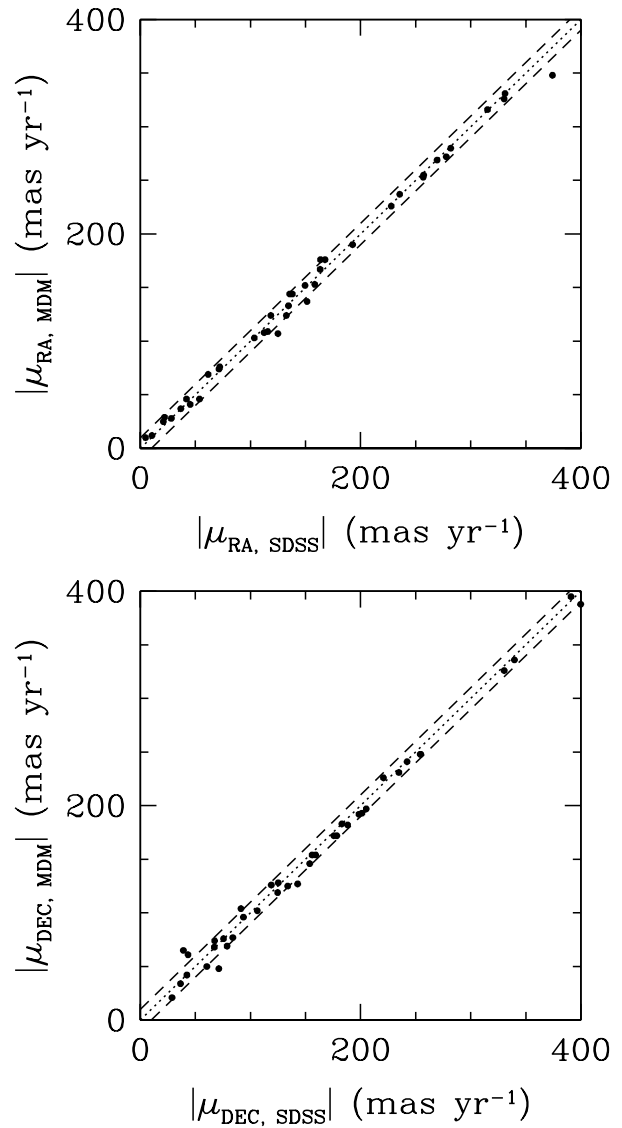}
\caption{Comparison between the proper motions measured at MDM
  Observatory and those from the SDSS+USNO-B catalogue \citep{munn04}
  for 42 of the 54 WDs in the current sample. We compare the absolute
  value of the proper motion in right ascension ($|\mu_{\rm RA}|$,
  top) and in declination ($|\mu_{\rm DEC}|$, bottom). The dotted line
  represents the 1:1 correlation.  The dashed lines represent the
  $\pm$~10~mas~yr$^{-1}$ range.}
\label{fg:pm}
\end{figure}

Our reduction and analysis procedures differed only slightly from
those described by \citet{thor03} and \citet{thor08}. As in the
previous work, we corrected our raw parallaxes to absolute using
colour-based distance estimates for the reference stars, and estimated
uncertainties using the formal errors of the fit and the scatter of
the references stars. In order to correct for differential colour
refraction (DCR), we need to know the colour of both the programme star
and the reference stars. In previous work we measured the colours, but
for this work we used SDSS $g - i$ colours and adjusted the DCR
correction factor slightly to account for this. \citet{thor03}
describes a Bayesian procedure used to estimate distances from the
available data, which combines the parallax measurement with an
assumed space velocity distribution and absolute magnitude range.  We
used a similar approach here, but modified the prior information to be
appropriate to the present sample. For the velocities, we used a
composite distribution consisting of 60 per cent thin disc with $(U,
V, W) = (39, 20, 20)\ {\rm km\ s^{-1}}$, 30 per cent thick disc with
$(U, V, W) = (46, 50, 35)\ {\rm km\ s^{-1}}$ \citep{chiba00}, and a 10
per cent probability of a still larger dispersion $(U, V, W) = (100,
75, 50)\ {\rm km\ s^{-1}}$. The absolute magnitudes of these WDs
are likely to be in the range 11--18, so the absolute
magnitudes were assumed to be drawn from a Gaussian centred on $M_{\rm g} =
15$ with a standard deviation of 4 mag. In most cases our
parallaxes were accurate enough that the Bayesian adjustments were
fairly minor. Furthermore, we have four targets in common with the USNO
Parallax programme and the parallax measurements are in good agreement
(Harris, private communication).

There is only one target in our parallax sample, J1547+0523
(NLTT~41210), that does not display significant parallax. This object
was identified as a high proper motion target by \citet{lepine05}, and
included in our sample as a WD candidate. We measure relative proper
motions of $-150.5 \pm 1.1$ and $-133.9 \pm 1.1$ mas yr$^{-1}$ in RA
and DEC, respectively. These are consistent with the proper motion
measurements by \citet{lepine05}. We also measure a parallax of 1.9
$\pm$ 1.4 mas, which indicates that NLTT~41210 is not a WD.

\subsection{Proper Motion}

In Fig.~\ref{fg:pm}, we compare our measured proper motions, as
listed in Table~\ref{tab:astro}, for the 42 WDs in our sample that
also have measured proper motions in the SDSS+USNO-B catalogue
\citep{munn04}. We expect disagreement at the 10~mas~yr$^{-1}$ level
since our proper motions are relative to the particular reference
stars used in the reduction. Fig.~\ref{fg:pm} shows that the vast
majority of our WDs do indeed fall within the range of
$\pm$~10~mas~yr$^{-1}$ when compared with the SDSS+USNO-B
measurements.

This disagreement arises due to two main factors. First, we make no
attempt to reduce proper motions to an inertial frame. Any systematic
trend due to e.g., Galactic rotation or solar motion, is still
present. Secondly, reference stars often have detectable proper motions
of their own, so in $\mu_{\rm RA}$~versus~$\mu_{\rm DEC}$ space they form
a cloud of points around the origin.  Because there are typically only
a couple of dozen reference stars, the centre of this cloud is
statistically uncertain, typically of the order of 5~mas~yr$^{-1}$.

\setlength{\extrarowheight}{1pt}
\setlength{\tabcolsep}{5.5pt}
\begin{table*}
\scriptsize
\centering
\begin{minipage}{0.99\textwidth}
\caption{Astrometry of cool WDs.}
\begin{tabular}{@{}lccr@{ $\pm$ }l*{2}{S[table-format=3.0]}r@{}lr@{ $\pm$ }lr@{ $\pm$ }lr@{ $\pm$ }lr@{ $\pm$ }l}
\hline
\noalign{\smallskip}
SDSS     & RA (J2000)   & Dec. (J2000) & \multicolumn{2}{c}{$\pi_{\rm abs}$}  & \multicolumn{1}{c}{$\mu_{\rm RA}$}   & \multicolumn{1}{c}{$\mu_{\rm DEC}$}   & \multicolumn{2}{c}{$D$}  & \multicolumn{2}{c}{$v_{\rm tan}$} & \multicolumn{2}{c}{$U^{a}$} & \multicolumn{2}{c}{$V$} & \multicolumn{2}{c}{$W$} \\
         & (h:m:s)      & (d:m:s)      & \multicolumn{2}{c}{(mas yr$^{-1}$)} & \multicolumn{1}{c}{(mas yr$^{-1}$)} & \multicolumn{1}{c}{(mas yr$^{-1}$)} & \multicolumn{2}{c}{(pc)} & \multicolumn{2}{c}{(km s$^{-1}$)} & \multicolumn{2}{c}{(km s$^{-1}$)} & \multicolumn{2}{c}{(km s$^{-1}$)} & \multicolumn{2}{c}{(km s$^{-1}$)} \\
\noalign{\smallskip}
\hline
\noalign{\smallskip}
J0045+1420   & 00:45:21.89 &   +14:20:45.3 & 15.9 & 1.1 &  251 &  -52 &  66 & $^{+6}_{-5}$   &  80 &  6 & $-$46 &  5 &  $-$37 &  4 &  $-$5 &  2 \\ 
J0121$-$0038 & 01:21:03.00 & $-$00:38:33.6 &  9.6 & 1.9 &  107 &   61 & 118 & $^{+29}_{-21}$ &  69 & 14 & $-$46 & 12 &      4 &  5 &    27 &  4 \\ 
J0146+1404   & 01:46:29.01 &   +14:04:38.2 & 11.0 & 1.5 &  255 &   65 &  90 & $^{+14}_{-10}$ & 112 & 14 & $-$82 & 13 &  $-$37 &  8 &    49 &  6 \\ 
J0256$-$0700 & 02:56:41.62 & $-$07:00:33.8 & 16.4 & 1.5 &  348 & -193 &  61 & $^{+6}_{-5}$   & 115 & 10 & $-$17 &  3 &  $-$96 & 10 &    34 &  3 \\ 
J0301$-$0044 & 03:01:44.09 & $-$00:44:39.5 & 13.4 & 1.3 &  107 & -538 &  74 & $^{+7}_{-6}$   & 192 & 16 &    83 &  7 & $-$155 & 17 & $-$60 &  7 \\ 
J0309+0025   & 03:09:24.87 &   +00:25:25.3 & 21.2 & 1.6 &  -10 & -102 &  49 & $^{+4}_{-4}$   &  24 &  1 &    23 &  2 &   $-$3 &  3 &  $-$4 &  2 \\ 
J0310$-$0110 & 03:10:49.53 & $-$01:10:35.3 &  7.1 & 1.9 &  -28 &  -77 & 164 & $^{+58}_{-35}$ &  64 & 18 &    47 & 11 &  $-$14 & 10 & $-$25 & 10 \\ 
J0747+2438N$^{b}$  & 07:47:21.56 &   +24:38:47.7 & 18.4 & 1.0 &  139 &  -69 &  55 & $^{+3}_{-3}$   &  40 &  2 &    29 &  2 &  $-$13 &  3 &    32 &  3 \\ 
J0747+2438S$^{b}$  & 07:47:23.50 &   +24:38:23.7 & 18.4 & 1.0 &  139 &  -69 &  55 & $^{+3}_{-3}$   &  40 &  2 &    29 &  2 &  $-$13 &  3 &    32 &  3 \\ 
J0753+4230   & 07:53:13.28 &   +42:30:01.6 & 36.2 & 1.0 &  108 & -388 &  27 & $^{+1}_{-1}$   &  53 &  1 &    10 &  1 &  $-$40 &  2 &    10 &  1 \\ 
J0805+3833   & 08:05:57.62 &   +38:33:44.1 & 47.6 & 1.0 & -628 & -552 &  21 & $^{+1}_{-1}$   &  83 &  1 & $-$24 &  1 &  $-$30 &  1 & $-$55 &  2 \\ 
J0817+2822   & 08:17:51.52 &   +28:22:03.1 & 19.6 & 1.5 &   69 & -197 &  52 & $^{+4}_{-4}$   &  51 &  3 &    25 &  2 &  $-$36 &  4 &     9 &  2 \\ 
J0821+3727   & 08:21:08.18 &   +37:27:38.3 & 12.8 & 3.2 &  167 & -146 &  86 & $^{+29}_{-19}$ &  90 & 25 &    44 &  8 &  $-$50 & 16 &    50 & 11 \\ 
J0825+5049   & 08:25:19.70 &   +50:49:20.1 & 20.4 & 1.3 & -331 & -326 &  49 & $^{+3}_{-3}$   & 108 &  6 & $-$48 &  4 &  $-$51 &  5 & $-$57 &  4 \\ 
J0854+3503   & 08:54:43.33 &   +35:03:52.7 & 15.7 & 1.5 & -144 & -172 &  64 & $^{+7}_{-6}$   &  68 &  6 & $-$14 &  3 &  $-$37 &  6 & $-$32 &  4 \\ 
J0909+4700   & 09:09:14.56 &   +47:00:17.5 & 16.6 & 1.7 & -124 & -172 &  61 & $^{+7}_{-6}$   &  61 &  6 & $-$17 &  3 &  $-$36 &  6 & $-$16 &  3 \\ 
J0942+4437   & 09:42:44.96 &   +44:37:43.1 & 11.7 & 1.2 & -144 & -182 &  87 & $^{+10}_{-8}$  &  96 &  9 & $-$30 &  5 &  $-$66 &  9 & $-$23 &  4 \\ 
J1001+3903   & 10:01:03.42 &   +39:03:40.5 & 11.6 & 2.1 & -291 & -175 &  87 & $^{+19}_{-13}$ & 140 & 25 & $-$73 & 16 &  $-$75 & 16 & $-$60 & 12 \\ 
J1107+4855   & 11:07:31.38 &   +48:55:23.0 & 20.9 & 1.7 & -730 &  -69 &  48 & $^{+4}_{-4}$   & 167 & 13 &   128 & 12 &  $-$56 &  6 & $-$53 &  5 \\ 
J1115+0033   & 11:15:36.97 &   +00:33:15.3 & 20.2 & 2.5 &   56 & -246 &  51 & $^{+7}_{-6}$   &  61 &  7 &    50 &  5 &  $-$27 &  5 & $-$14 &  3 \\ 
J1117+5010   & 11:17:08.63 &   +50:10:33.9 & 21.3 & 1.7 &  176 & -126 &  48 & $^{+4}_{-4}$   &  49 &  4 &    51 &  4 &      1 &  2 &    31 &  2 \\ 
J1158+0004   & 11:58:14.52 &   +00:04:58.3 & 28.9 & 1.7 &  -25 &  183 &  35 & $^{+2}_{-2}$   &  31 &  1 &  $-$7 &  2 &     33 &  2 &    20 &  1 \\ 
J1203+0426   & 12:03:28.64 &   +04:26:53.6 & 22.8 & 2.1 & -253 &  154 &  45 & $^{+5}_{-4}$   &  63 &  6 & $-$50 &  6 &     13 &  2 &    10 &  1 \\ 
J1204+6222   & 12:04:39.54 &   +62:22:16.4 & 18.3 & 2.5 &  -29 & -154 &  58 & $^{+10}_{-7}$  &  43 &  6 &    16 &  2 &  $-$22 &  5 &    28 &  3 \\ 
J1212+0440   & 12:12:07.01 &   +04:40:12.0 & 16.0 & 2.6 & -280 &  -48 &  64 & $^{+13}_{-9}$  &  86 & 14 & $-$54 & 11 &  $-$38 &  9 & $-$11 &  3 \\ 
J1238+3502   & 12:38:12.85 &   +35:02:49.1 & 10.0 & 2.0 & -146 & -123 & 110 & $^{+26}_{-19}$ & 100 & 20 & $-$18 &  7 &  $-$73 & 18 &    12 &  1 \\ 
J1251+4403   & 12:51:06.12 &   +44:03:03.1 & 22.9 & 5.3 &   22 & -136 &  63 & $^{+29}_{-16}$ &  41 & 14 &    30 &  5 &   $-$8 &  5 &    15 &  2 \\ 
J1345+4200   & 13:45:32.92 &   +42:00:44.2 & 27.3 & 1.0 & -190 &  128 &  36 & $^{+1}_{-1}$   &  40 &  1 & $-$28 &  2 &      8 &  2 &    9 &  1 \\ 
J1349+1155   & 13:49:02.33 &   +11:55:11.8 & 35.3 & 1.6 &  176 & -524 &  28 & $^{+1}_{-1}$   &  74 &  3 &    69 &  3 &  $-$28 &  2 & $-$17 &  1 \\ 
J1422+0459   & 14:22:25.73 &   +04:59:39.7 & 16.7 & 2.1 & -272 &  -50 &  61 & $^{+9}_{-7}$   &  80 & 10 & $-$34 &  6 &  $-$49 &  8 &    28 &  3 \\ 
J1424+6246   & 14:24:29.52 &   +62:46:17.1 & 21.1 & 2.0 & -269 &  -42 &  48 & $^{+5}_{-4}$   &  62 &  5 & $-$23 &  4 &  $-$31 &  5 &    33 &  3 \\ 
J1436+4332   & 14:36:42.78 &   +43:32:35.7 & 37.1 & 1.2 & -316 &  498 &  27 & $^{+1}_{-1}$   &  75 &  2 & $-$63 &  3 &     25 &  1 &     9 &  1 \\ 
J1437+4151   & 14:37:18.15 &   +41:51:51.5 & 16.0 & 2.2 & -153 &  -68 &  66 & $^{+12}_{-8}$  &  52 &  7 &  $-$3 &  3 &  $-$30 &  6 &    29 &  3 \\ 
J1447+5427   & 14:47:01.85 &   +54:27:44.6 & 21.3 & 3.5 & -237 &   34 &  51 & $^{+11}_{-8}$  &  58 & 10 & $-$27 &  7 &  $-$19 &  5 &    27 &  4 \\ 
J1452+4522   & 14:52:39.00 &   +45:22:38.3 & 11.4 & 1.4 &  -46 &   74 &  95 & $^{+15}_{-11}$ &  39 &  5 & $-$25 &  6 &     17 &  4 &     9 &  2 \\ 
J1458+1146   & 14:58:48.52 &   +11:46:55.9 & 17.0 & 1.6 & -124 &  -96 &  60 & $^{+6}_{-5}$   &  45 &  4 &     7 &  2 &  $-$31 &  5 &    16 &  2 \\ 
J1534+4649   & 15:34:51.02 &   +46:49:49.5 & 33.1 & 1.8 & -468 &  226 &  30 & $^{+1}_{-1}$   &  75 &  4 & $-$49 &  4 &  $-$17 &  2 &    40 &  2 \\ 
J1606+2547   & 16:06:19.81 &   +25:47:02.9 & 22.6 & 1.7 & -226 & -119 &  45 & $^{+4}_{-3}$   &  54 &  4 &    13 &  2 &  $-$33 &  4 &    35 &  3 \\ 
J1615+4449   & 16:15:44.67 &   +44:49:42.5 & 12.1 & 3.7 &   46 & -231 &  89 & $^{+47}_{-20}$ &  99 & 37 &   100 & 27 &  $-$12 &  8 &     1 &  3 \\ 
J1632+2426   & 16:32:42.23 &   +24:26:55.2 & 22.9 & 1.2 &  -12 & -336 &  44 & $^{+2}_{-2}$   &  70 &  3 &    65 &  3 &  $-$30 &  3 &  $-$5 &  2 \\ 
J1704+3608   & 17:04:47.70 &   +36:08:47.4 & 21.1 & 1.7 &  183 & -172 &  48 & $^{+4}_{-4}$   &  57 &  4 &    52 &  4 &     20 &  2 & $-$31 &  4 \\ 
J1722+5752   & 17:22:57.78 &   +57:52:50.7 & 17.8 & 1.8 &  -37 &  395 &  56 & $^{+6}_{-5}$   & 105 & 10 & $-$94 & 11 &     16 &  2 &    11 &  2 \\ 
J1728+2646   & 17:28:07.27 &   +26:46:19.2 & 17.2 & 1.8 &  -41 & -248 &  59 & $^{+7}_{-6}$   &  70 &  7 &    67 &  6 &  $-$28 &  5 &  $-$1 &  3 \\ 
J2041$-$0520 & 20:41:28.99 & $-$05:20:27.7 & 15.8 & 1.6 & -152 &  -21 &  65 & $^{+8}_{-6}$   &  47 &  5 &    41 &  4 &      3 &  3 &    41 &  4 \\ 
J2042+0031   & 20:42:59.23 &   +00:31:56.6 & 16.0 & 1.3 &  -76 & -241 &  63 & $^{+6}_{-5}$   &  75 &  6 &    60 &  5 &  $-$42 &  5 &  $-$8 &  3 \\ 
J2045$-$0710 & 20:45:57.53 & $-$07:10:03.5 & 12.0 & 1.4 &  -74 & -125 &  86 & $^{+12}_{-9}$  &  59 &  7 &    49 &  5 &  $-$31 &  6 &    10 &  3 \\ 
J2118$-$0737 & 21:18:05.21 & $-$07:37:29.1 & 14.5 & 2.0 &  109 & -127 &  72 & $^{+12}_{-9}$  &  57 &  8 &     3 &  3 &  $-$23 &  6 & $-$34 &  6 \\ 
J2147+1127   & 21:47:25.17 &   +11:27:56.1 & 18.9 & 2.0 &  103 & -248 &  54 & $^{+7}_{-5}$   &  69 &  7 &    26 &  3 &  $-$28 &  5 & $-$45 &  6 \\ 
J2222+1221   & 22:22:33.89 &   +12:21:43.0 & 24.4 & 1.3 &  703 &  192 &  41 & $^{+2}_{-2}$   & 142 &  6 &   121 &  7 &      7 &  1 & $-$44 &  3 \\ 
J2239+0018A$^{b}$  & 22:39:54.12 &   +00:18:47.3 & 12.2 & 2.9 &  -12 &  122 & 107 & $^{+42}_{-26}$ &  62 & 19 &  $-$8 &  6 &     49 &  9 &    31 &  6 \\ 
J2239+0018B$^{b}$  & 22:39:54.07 &   +00:18:49.2 & 12.2 & 2.9 &  -12 &  122 & 107 & $^{+42}_{-26}$ &  62 & 19 &  $-$8 &  6 &     49 &  9 &    31 &  6 \\ 
J2242+0048   & 22:42:06.19 &   +00:48:22.8 & 14.8 & 1.7 &  133 &  -76 &  70 & $^{+9}_{-7}$   &  51 &  5 & $-$13 &  4 &  $-$17 &  4 & $-$24 &  4 \\ 
J2254+1323   & 22:54:08.64 &   +13:23:57.2 & 24.2 & 1.6 &  326 & -192 &  42 & $^{+3}_{-3}$   &  75 &  5 & $-$24 &  3 &  $-$28 &  3 & $-$44 &  4 \\ 
J2330+0028   & 23:30:55.20 &   +00:28:52.3 & 18.3 & 2.3 &  137 &  104 &  59 & $^{+9}_{-7}$   &  48 &  6 & $-$33 &  6 &     19 &  2 &     9 &  1 \\ 
\noalign{\smallskip}
\hline
\noalign{\smallskip}
\multicolumn{17}{@{}p{\textwidth}@{}}{$^{a}$ Since we do not have
  any radial velocity measurements for our targets, the $U$ component
  has been computed assuming $v_{\rm rad}$~=~0~\kms.} \\
\multicolumn{17}{@{}p{\textwidth}@{}}{$^{b}$ For these two binary 
  systems, a weighted mean was adopted in the determination of their 
  astrometric measurements.}\\
\label{tab:astro}
\end{tabular}
\end{minipage}
\end{table*}

\subsection{Optical and Infrared Photometry}
We have obtained the available $ugriz$ photometry from the SDSS Data 
Release 10 \citep[DR10,][]{ahn14} for the 54 WDs in our sample. 
These data are listed in columns two through six in Table~\ref{tab:phot} 
along with their uncertainties. The majority of our targets also have 
near-infrared photometry available from \citet{kilic10a}, and are also 
listed in Table~\ref{tab:phot}. For the six WDs without near-infrared 
photometry from \citet{kilic10a}, we adopt the near-infrared photometry 
from the UKIDSS Large Area Survey (ULAS) Catalog \citep{lawrence07}, 
and the Two Micron All Sky Survey \citep[2MASS;][]{skrut06}; see the 
notes at the bottom of Table~\ref{tab:phot}.

\subsection{Optical Spectroscopy}
The majority of our targets were selected from the cool WD samples of
\citet{kilic06a,kilic10a}, hence they have optical
spectroscopy obtained at the McDonald Observatory 2.7m telescope,
Hobby-Eberly Telescope, or the Multiple-Mirror Telescope. The ultracool 
WDs and a few other cool WDs have spectroscopy available in the SDSS or 
the literature \citep{leggett11,giam12,tremblay14}. There are only eight 
DA WDs in our sample, with the rest of the stars classified as DC due to 
the absence of \halpha\ absorption. This overabundance of DC WDs is due 
to our selection bias for targeting cool and ultracool WDs.

\setlength{\extrarowheight}{0pt}
\setlength{\tabcolsep}{5pt}
\begin{table*}
\scriptsize
\centering
\begin{minipage}{\textwidth}
\caption{Optical and near-infrared photometry of cool WDs.}
\begin{tabular}{@{}lr@{ $\pm$ }lr@{ $\pm$ }lr@{ $\pm$ }lr@{ $\pm$ }lr@{ $\pm$ }lr@{ $\pm$ }lr@{ $\pm$ }lr@{ $\pm$ }lr@{ $\pm$ }l@{}}
\hline
\noalign{\smallskip}
SDSS     & \multicolumn{2}{c}{$u$} & \multicolumn{2}{c}{$g$} & \multicolumn{2}{c}{$r$} & \multicolumn{2}{c}{$i$} & \multicolumn{2}{c}{$z$} & \multicolumn{2}{c}{$Y$} & \multicolumn{2}{c}{$J$} & \multicolumn{2}{c}{$H$} & \multicolumn{2}{c}{$K$} \\
\noalign{\smallskip}
\hline
\noalign{\smallskip}
J0045+1420   & 20.64 & 0.08 & 19.20 & 0.03 & 18.45 & 0.03 & 18.20 & 0.03 & 18.10 & 0.03 & \multicolumn{2}{c}{--} & 17.24 & 0.04 & 16.99 & 0.04 & 16.89 & 0.04 \\
J0121$-$0038$^{a}$ & 22.82 & 0.28 & 20.79 & 0.03 & 19.74 & 0.03 & 19.38 & 0.03 & 19.18 & 0.04 &    18.47     &     0.06    & 18.23 & 0.08 & 18.05 & 0.09 & 18.10 & 0.19 \\
J0146+1404   & 21.37 & 0.11 & 20.00 & 0.03 & 19.39 & 0.02 & 19.27 & 0.03 & 19.79 & 0.11 & \multicolumn{2}{c}{--} & 19.56 & 0.05 & 20.07 & 0.12 & \multicolumn{2}{c}{--} \\
J0256$-$0700 & 20.74 & 0.08 & 19.00 & 0.02 & 18.13 & 0.02 & 17.79 & 0.03 & 17.69 & 0.03 & \multicolumn{2}{c}{--} & 16.71 & 0.05 & 16.62 & 0.05 & 16.48 & 0.06 \\
J0301$-$0044 & 22.23 & 0.34 & 20.43 & 0.03 & 19.38 & 0.02 & 18.99 & 0.02 & 18.92 & 0.04 & \multicolumn{2}{c}{--} & 17.96 & 0.04 & 17.73 & 0.04 & 17.68 & 0.08 \\
J0309+0025   & 19.15 & 0.03 & 18.19 & 0.02 & 17.72 & 0.02 & 17.53 & 0.02 & 17.50 & 0.02 & \multicolumn{2}{c}{--} & 16.64 & 0.04 & 16.54 & 0.04 & 16.87 & 0.04 \\
J0310$-$0110 & 22.71 & 0.30 & 20.89 & 0.04 & 20.18 & 0.03 & 19.91 & 0.03 & 19.75 & 0.08 & \multicolumn{2}{c}{--} & 18.94 & 0.02 & 18.73 & 0.02 & 18.58 & 0.02 \\
J0747+2438N  & 21.01 & 0.08 & 19.29 & 0.02 & 18.59 & 0.02 & 18.23 & 0.01 & 18.14 & 0.02 & \multicolumn{2}{c}{--} & 17.16 & 0.04 & 16.99 & 0.04 & 16.85 & 0.04 \\
J0747+2438S  & 19.49 & 0.03 & 18.37 & 0.01 & 17.91 & 0.01 & 17.73 & 0.01 & 17.69 & 0.02 & \multicolumn{2}{c}{--} & 16.78 & 0.04 & 16.58 & 0.04 & 16.53 & 0.04 \\
J0753+4230   & 19.97 & 0.04 & 18.09 & 0.01 & 17.19 & 0.01 & 16.87 & 0.01 & 16.75 & 0.02 & \multicolumn{2}{c}{--} & 15.69 & 0.04 & 15.49 & 0.04 & 15.47 & 0.04 \\
J0804+2239   & 19.73 & 0.03 & 18.30 & 0.02 & 17.59 & 0.01 & 17.39 & 0.01 & 17.33 & 0.02 & \multicolumn{2}{c}{--} & 16.71 & 0.04 & 16.92 & 0.04 & 17.29 & 0.06 \\
J0805+3833$^{b}$ & 19.00 & 0.02 & 17.31 & 0.01 & 16.56 & 0.02 & 16.27 & 0.02 & 16.20 & 0.02 & \multicolumn{2}{c}{--} & 15.34 & 0.05 & 15.19 & 0.08 & 14.90 & 0.09 \\
J0817+2822   & 21.59 & 0.16 & 19.49 & 0.02 & 18.61 & 0.01 & 18.30 & 0.01 & 18.22 & 0.03 & \multicolumn{2}{c}{--} & 17.33 & 0.04 & 17.01 & 0.04 & 16.91 & 0.09 \\
J0821+3727   & 20.68 & 0.06 & 19.14 & 0.02 & 18.43 & 0.01 & 18.15 & 0.02 & 18.04 & 0.02 & \multicolumn{2}{c}{--} & 17.25 & 0.04 & 17.00 & 0.04 & 16.85 & 0.05 \\
J0825+5049   & 21.09 & 0.09 & 19.34 & 0.02 & 18.43 & 0.02 & 18.09 & 0.02 & 18.00 & 0.03 & \multicolumn{2}{c}{--} & 17.08 & 0.04 & 16.83 & 0.04 & 16.74 & 0.04 \\
J0845+2257   & 15.57 & 0.01 & 15.73 & 0.01 & 16.08 & 0.01 & 16.35 & 0.02 & 16.61 & 0.02 & \multicolumn{2}{c}{--} & 16.24 & 0.11 & 15.96 & 0.00 & 16.48 & 0.00 \\
J0854+3503   & 23.57 & 0.67 & 20.53 & 0.03 & 19.39 & 0.02 & 19.09 & 0.03 & 18.95 & 0.05 & \multicolumn{2}{c}{--} & 18.44 & 0.04 & 18.23 & 0.04 & 17.98 & 0.04 \\
J0909+4700   & 20.64 & 0.15 & 19.29 & 0.03 & 18.74 & 0.02 & 18.50 & 0.02 & 18.42 & 0.05 & \multicolumn{2}{c}{--} & 18.11 & 0.04 & 18.62 & 0.07 & 19.10 & 0.10 \\
J0942+4437   & 21.37 & 0.09 & 19.47 & 0.02 & 18.58 & 0.01 & 18.22 & 0.02 & 18.05 & 0.02 & \multicolumn{2}{c}{--} & 17.15 & 0.04 & 16.97 & 0.04 & 16.86 & 0.04 \\
J1001+3903   & 21.36 & 0.10 & 20.05 & 0.02 & 19.60 & 0.02 & 20.02 & 0.03 & 20.61 & 0.17 & \multicolumn{2}{c}{--} & 20.65 & 0.06 & 21.05 & 0.07 & \multicolumn{2}{c}{--} \\
J1107+4855   & 21.50 & 0.12 & 19.49 & 0.03 & 18.54 & 0.02 & 18.23 & 0.02 & 18.11 & 0.03 & \multicolumn{2}{c}{--} & 17.05 & 0.05 & 16.95 & 0.07 & 16.86 & 0.07 \\
J1115+0033$^{a}$ & 19.50 & 0.04 & 17.92 & 0.01 & 17.22 & 0.02 & 16.99 & 0.01 & 16.90 & 0.02 & \multicolumn{2}{c}{--} & 15.78 & 0.08 & 15.65 & 0.18 & 15.59 & 0.26 \\
J1117+5010   & 21.17 & 0.10 & 19.34 & 0.03 & 18.57 & 0.03 & 18.30 & 0.02 & 18.16 & 0.03 & \multicolumn{2}{c}{--} & 17.24 & 0.04 & 17.07 & 0.04 & 16.97 & 0.05 \\
J1158+0004   & 20.86 & 0.11 & 18.89 & 0.04 & 17.85 & 0.02 & 17.54 & 0.01 & 17.34 & 0.03 & \multicolumn{2}{c}{--} & 16.36 & 0.04 & 16.31 & 0.05 & 16.18 & 0.05 \\
J1203+0426   & 19.57 & 0.03 & 18.18 & 0.02 & 17.50 & 0.02 & 17.21 & 0.01 & 17.12 & 0.02 & \multicolumn{2}{c}{--} & 16.39 & 0.01 & 16.49 & 0.02 & 16.92 & 0.06 \\
J1204+6222   & 20.91 & 0.09 & 19.25 & 0.02 & 18.43 & 0.02 & 18.14 & 0.02 & 18.06 & 0.03 & \multicolumn{2}{c}{--} & 17.07 & 0.04 & 16.86 & 0.04 & 16.80 & 0.04 \\
J1212+0440   & 22.07 & 0.21 & 20.04 & 0.03 & 19.09 & 0.02 & 18.79 & 0.02 & 18.66 & 0.04 & \multicolumn{2}{c}{--} & 17.67 & 0.04 & 17.50 & 0.04 & 17.50 & 0.05 \\
J1238+3502   & 24.74 & 0.81 & 21.77 & 0.09 & 20.31 & 0.06 & 19.88 & 0.05 & 20.37 & 0.15 & \multicolumn{2}{c}{--} & 21.19 & 0.06 & \multicolumn{2}{c}{--} & \multicolumn{2}{c}{--} \\
J1251+4403   & 21.46 & 0.09 & 20.17 & 0.03 & 20.39 & 0.03 & 20.72 & 0.04 & 20.92 & 0.17 & \multicolumn{2}{c}{--} & 21.78 & 0.08 & \multicolumn{2}{c}{--} & \multicolumn{2}{c}{--} \\
J1345+4200$^{b}$ & 19.70 & 0.03 & 17.85 & 0.02 & 17.01 & 0.01 & 16.69 & 0.02 & 16.54 & 0.01 & \multicolumn{2}{c}{--} & 15.61 & 0.06 & 15.43 & 0.11 & 15.00 & 0.00 \\
J1349+1155   & 20.55 & 0.06 & 18.64 & 0.02 & 17.84 & 0.01 & 17.42 & 0.02 & 17.20 & 0.02 & \multicolumn{2}{c}{--} & 16.43 & 0.01 & 16.29 & 0.02 & 16.26 & 0.02 \\
J1422+0459   & 20.98 & 0.10 & 19.44 & 0.03 & 18.58 & 0.02 & 18.27 & 0.02 & 18.18 & 0.03 & \multicolumn{2}{c}{--} & 17.15 & 0.05 & 17.10 & 0.08 & 17.02 & 0.05 \\
J1424+6246   & 20.38 & 0.05 & 18.83 & 0.01 & 18.15 & 0.03 & 17.89 & 0.02 & 17.71 & 0.02 & \multicolumn{2}{c}{--} & \multicolumn{2}{c}{--} & \multicolumn{2}{c}{--} & \multicolumn{2}{c}{--} \\
J1436+4332   & 19.83 & 0.04 & 18.04 & 0.02 & 17.19 & 0.01 & 16.85 & 0.03 & 16.75 & 0.03 & \multicolumn{2}{c}{--} & 15.78 & 0.04 & 15.62 & 0.04 & 15.51 & 0.04 \\
J1437+4151   & 20.06 & 0.04 & 19.03 & 0.01 & 18.45 & 0.02 & 18.23 & 0.03 & 18.12 & 0.02 & \multicolumn{2}{c}{--} & 17.43 & 0.04 & 17.76 & 0.05 & 18.42 & 0.08 \\
J1447+5427   & 21.23 & 0.12 & 19.46 & 0.04 & 18.64 & 0.02 & 18.36 & 0.02 & 18.25 & 0.04 & \multicolumn{2}{c}{--} & 17.26 & 0.07 & 17.20 & 0.07 & 17.07 & 0.06 \\
J1452+4522   & 21.55 & 0.10 & 20.01 & 0.02 & 19.39 & 0.02 & 19.31 & 0.02 & 19.36 & 0.06 & \multicolumn{2}{c}{--} & 18.60 & 0.02 & 18.43 & 0.02 & 18.37 & 0.02 \\
J1458+1146   & 20.62 & 0.08 & 18.85 & 0.02 & 18.02 & 0.02 & 17.72 & 0.02 & 17.64 & 0.02 & \multicolumn{2}{c}{--} & 16.63 & 0.04 & 16.47 & 0.05 & 16.31 & 0.06 \\
J1534+4649   & 20.90 & 0.08 & 18.76 & 0.02 & 17.74 & 0.02 & 17.36 & 0.02 & 17.19 & 0.02 & \multicolumn{2}{c}{--} & 16.17 & 0.04 & 16.12 & 0.04 & 16.04 & 0.05 \\
J1547+0523   & 19.96 & 0.04 & 18.05 & 0.01 & 17.13 & 0.00 & 16.75 & 0.00 & 16.51 & 0.01 & \multicolumn{2}{c}{--} & 15.38 & 0.00 & 14.95 & 0.00 & 14.77 & 0.01 \\
J1606+2547   & 20.99 & 0.08 & 19.24 & 0.02 & 18.45 & 0.02 & 18.17 & 0.02 & 18.07 & 0.04 & \multicolumn{2}{c}{--} & 17.07 & 0.04 & 17.09 & 0.06 & 16.84 & 0.06 \\
J1615+4449   & 21.18 & 0.10 & 19.59 & 0.02 & 18.84 & 0.02 & 18.57 & 0.02 & 18.52 & 0.04 & \multicolumn{2}{c}{--} & 17.44 & 0.04 & 17.24 & 0.05 & 17.26 & 0.07 \\
J1632+2426   & 21.47 & 0.10 & 19.60 & 0.02 & 18.73 & 0.02 & 18.49 & 0.02 & 18.40 & 0.03 & \multicolumn{2}{c}{--} & 17.67 & 0.02 & 18.10 & 0.02 & 18.04 & 0.02 \\
J1704+3608   & 20.50 & 0.05 & 18.72 & 0.01 & 17.94 & 0.01 & 17.66 & 0.01 & 17.55 & 0.02 & \multicolumn{2}{c}{--} & 16.62 & 0.04 & 16.34 & 0.04 & 16.32 & 0.06 \\
J1722+5752   & 20.39 & 0.06 & 19.28 & 0.02 & 18.79 & 0.02 & 18.56 & 0.03 & 18.50 & 0.03 & \multicolumn{2}{c}{--} & 17.74 & 0.04 & 17.84 & 0.05 & 18.75 & 0.12 \\
J1728+2646   & 19.18 & 0.03 & 18.14 & 0.02 & 17.68 & 0.01 & 17.51 & 0.01 & 17.46 & 0.02 & \multicolumn{2}{c}{--} & \multicolumn{2}{c}{--} & \multicolumn{2}{c}{--} & \multicolumn{2}{c}{--} \\
J2041$-$0520 & 20.95 & 0.08 & 19.27 & 0.01 & 18.51 & 0.01 & 18.24 & 0.01 & 18.14 & 0.03 & \multicolumn{2}{c}{--} & 17.25 & 0.04 & 16.97 & 0.04 & 16.93 & 0.04 \\
J2042+0031   & 21.67 & 0.14 & 19.95 & 0.02 & 19.05 & 0.01 & 18.73 & 0.01 & 18.61 & 0.03 & \multicolumn{2}{c}{--} & 17.65 & 0.04 & 17.45 & 0.04 & 17.36 & 0.05 \\
J2045$-$0710 & 21.03 & 0.09 & 19.33 & 0.02 & 18.60 & 0.01 & 18.33 & 0.01 & 18.15 & 0.03 & \multicolumn{2}{c}{--} & 17.32 & 0.04 & 17.10 & 0.04 & 17.03 & 0.04 \\
J2118$-$0737 & 23.38 & 0.95 & 20.70 & 0.03 & 19.48 & 0.03 & 19.01 & 0.02 & 18.76 & 0.04 & \multicolumn{2}{c}{--} & 17.90 & 0.04 & 17.82 & 0.04 & 17.81 & 0.05 \\
J2147+1127   & 20.83 & 0.09 & 19.19 & 0.02 & 18.43 & 0.02 & 18.13 & 0.02 & 18.01 & 0.03 & \multicolumn{2}{c}{--} & 17.14 & 0.04 & 16.84 & 0.04 & 16.79 & 0.04 \\
J2222+1221   & 21.74 & 0.11 & 19.48 & 0.02 & 18.37 & 0.02 & 17.88 & 0.02 & 17.67 & 0.02 & \multicolumn{2}{c}{--} &  \multicolumn{2}{c}{--} & \multicolumn{2}{c}{--} & \multicolumn{2}{c}{--} \\
J2239+0018A$^{a}$ & 21.27 & 0.08 & 20.14 & 0.04 & 19.59 & 0.03 & 19.48 & 0.05 & 20.28 & 0.16 & 19.57 & 0.10 & 19.69       & 0.19 & \multicolumn{2}{c}{--} & \multicolumn{2}{c}{--} \\
J2239+0018B$^{a}$ & 23.13 & 0.36 & 20.79 & 0.04 & 19.88 & 0.03 & 19.49 & 0.03 & 19.24 & 0.06 & 18.66 & 0.05 & 18.34       & 0.06 & 17.98 & 0.10 & 18.48 & 0.27 \\
J2242+0048   & 22.11 & 0.22 & 19.63 & 0.02 & 18.65 & 0.02 & 18.28 & 0.01 & 18.16 & 0.03 & \multicolumn{2}{c}{--} & 18.06 & 0.04 & 18.72 & 0.07 & 19.16 & 0.10 \\
J2254+1323   & 21.57 & 0.17 & 19.51 & 0.02 & 18.49 & 0.02 & 18.14 & 0.01 & 18.00 & 0.02 & \multicolumn{2}{c}{--} & 17.04 & 0.04 & 16.88 & 0.04 & 16.85 & 0.04 \\
J2330+0028   & 21.85 & 0.24 & 19.88 & 0.02 & 18.95 & 0.02 & 18.66 & 0.02 & 18.53 & 0.04 & \multicolumn{2}{c}{--} & 17.63 & 0.04 & 17.36 & 0.04 & 17.32 & 0.04 \\
\noalign{\smallskip}
\hline
\noalign{\smallskip}
\multicolumn{19}{@{}p{\textwidth}@{}}{$^{a}$ IR photometry from UKIDSS} \\
\multicolumn{19}{@{}p{\textwidth}@{}}{$^{b}$ IR photometry from 2MASS} \\
\label{tab:phot}
\end{tabular}
\end{minipage}
\end{table*}

\section{THEORETICAL FRAMEWORK}
Our model atmospheres and synthetic spectra are derived from the local 
thermodynamic equilibrium (LTE) model atmosphere code originally 
described in \citet{bergeron95} and references therein, with recent 
improvements discussed in \citet{tb09}. In particular, we now rely on 
their improved calculations for the Stark broadening of hydrogen lines 
with the inclusion of non-ideal perturbations from protons and electrons 
-- described within the occupation probability formalism of \citet{hm88}
-- directly inside the line profile calculations. Convective energy
transport is taken into account following the ML2/$\alpha$~=~0.7
prescription of the mixing length theory. Non-LTE effects are also
included at higher effective temperatures but these are irrelevant for
the purpose of this work. More details regarding our helium-atmosphere
models are provided in \citet{bergeron11}.

Our model grid covers a range of effective temperature between
\Te\ =~1500 and 45,000~K in steps of 500~K for \Te\ $<$~15,000~K,
1000~K up to \Te\ =~18,000~K, 2000~K up to \Te\ =~30,000~K and by
steps of 5000~K above. The \logg\ ranges from 6.5 to 9.5 by steps of
0.5 dex, with additional models at \logg\ =~7.75 and 8.25.  We also
calculated mixed hydrogen and helium atmosphere models with
\loghe\ =~$-$2.0 to 5.0, in steps of 1.0 dex.

Since the photometric technique described below relies heavily on the
flux at the $u$ and $g$ bandpasses, we now include in our models the
opacity from the red wing of Ly$\alpha$ \citep{kowalski06}, which
significantly affects the flux in the ultraviolet.

\section{PHOTOMETRIC ANALYSIS}

\subsection{General Procedure}

\begin{figure*}
\centering
\includegraphics[scale=0.475,angle=-90,bb=22 16 606 784]{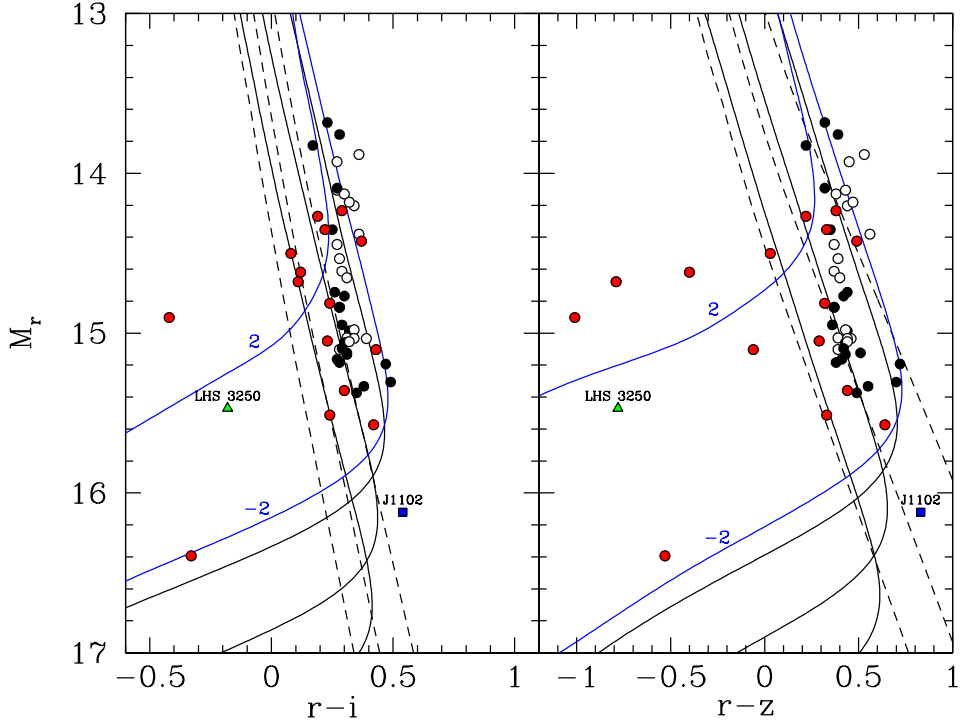}
\caption{Location of the WDs in our sample in a $M_{r}$ versus $r-i$
  (left) and $r-z$ (right) colour-magnitude diagram. The black dots
  correspond to the WDs with H-rich atmospheres while the white dots
  represent the He-rich WDs. The red dots represent the 15 cool and
  ultracool WDs with mixed atmospheres. The solid and dashed black
  lines represent pure H and pure He tracks, respectively, for masses
  $M$ =~0.3, 0.6, and 0.9 \msun, from right to left. The solid blue
  lines represent the predictions from mixed model atmospheres for $M$
  = 0.2~\msun. The mixed atmosphere model tracks are labelled with
  their He abundance \loghe\ = $-$2 and 2. LHS~3250 and J1102 are
  shown as a green triangle, and a blue square, respectively.
\label{fg:color}}
\end{figure*}

Atmospheric parameters, \Te\ and \logg, and chemical compositions of
cool WDs can be measured accurately using the photometric
technique developed by \citet{brl97}. We first convert optical and
infrared photometric measurements into observed fluxes and compare the
resulting energy distributions with those predicted from our model
atmosphere calculations. To accomplish this task, we first transform
every magnitude $m$ into an average flux $f^{m}_{\lambda}$. Since
$ugriz$ photometry is defined on the AB magnitude system, we first
calculate $f^{m}_{\nu}$ using the equation

\begin{equation}
m = -2.5\log f^{m}_{\nu} - 48.60
\end{equation}

\noindent and then $f^{m}_{\nu}$ is converted to $f^{m}_{\lambda}$
following $f^{m}_{\lambda} = f^{m}_{\nu}c/\lambda^{2}$, where
$\lambda$ is the central wavelength of the given filter. For the
near-infrared photometry, we obtain $f^{m}_{\lambda}$ using the
equation

\begin{equation}
m = -2.5\log f^{m}_{\lambda} + c_{m},
\label{eq:mag}
\end{equation} 

\noindent where $c_{m}$ is a constant to be determined for each
filter, as described below. In general,

\begin{equation}
f^{m}_{\lambda} = \frac{\int^{\infty}_{0}f_{\lambda}S_{m}(\lambda)\lambda {\rm d}\lambda}{\int^{\infty}_{0}S_{m}(\lambda)\lambda {\rm d}\lambda}
\label{eq:avg}
\end{equation}

\noindent where $S_{m}(\lambda)$ is the transmission function of the
corresponding bandpass, $f_{\lambda}$ is the monochromatic flux from
the star received at Earth. For the $ugriz$ photometry, a slightly
different definition of the above Equation~(\ref{eq:avg}) is required
\citep[see Equation (3) of][for instance]{hb06}. The transmission
functions for the $ugriz$ system are described in \citet{hb06} and
references therein. The transmission functions for the $JHK$ or
$JHK_{S}$ filters on the MKO photometric system are taken from
\citet{tokunaga02}.

The constants $c_{m}$ in Equation (\ref{eq:mag}) for each passband are
determined using the improved calibration fluxes from \citet{hb06},
defined with the $HST$ absolute flux scale of Vega \citep{bohlin04}, 
and appropriate magnitudes on a given system.

For each star in Table~\ref{tab:phot}, a minimum set of five average
fluxes $f^{m}_{\lambda}$ is obtained, which can be compared with model
fluxes. Since the observed fluxes correspond to averages over given
bandpasses, the monochromatic fluxes from the model atmospheres need
to be converted into average fluxes, $H^{m}_{\lambda}$, by
substituting $f_{\lambda}$ in Equation (\ref{eq:avg}) for the
monochromatic Eddington flux, $H_{\lambda}$.  We can then relate the
average observed fluxes $f^{m}_{\lambda}$ and the average model fluxes
$H^{m}_{\lambda}$ -- which depend on \Te, \logg and chemical
composition -- by the equation

\begin{equation}
f^{m}_{\lambda} = 4{\pi} (R/D)^{2} H^{m}_{\lambda}
\end{equation}

\noindent where $R/D$ defines the ratio of the radius of the star to
its distance from Earth. We then minimize the $\chi^{2}$ value defined
in terms of the difference between observed and model fluxes over all
bandpasses, properly weighted by the photometric uncertainties. Our
minimization procedure relies on the non-linear least-squares method of
Levenberg--Marquardt \citep{press86}, which is based on a steepest
decent method. Only \Te\ and the solid angle $\pi(R/D)^{2}$ are
considered free parameters, while the uncertainties of both parameters
are obtained directly from the covariance matrix of the fit.

For stars with known trigonometric parallax measurements, we first
assume a value of \logg\ =~8.0 and determine the effective temperature
and the solid angle, which combined with the distance $D$ obtained
from the trigonometric parallax measurement, yields directly the
radius of the star $R$. The radius is then converted into mass using
evolutionary models similar to those described in \citet{fontaine01}
but with CO cores, $q$(He) $\equiv \log M_{\rm He}/M_{\star} =
10^{-2}$ and $q$(H)~=~10$^{-4}$, which are representative of
hydrogen-atmosphere WDs, and $q$(He)~=~10$^{-2}$ and
$q$(H)~=~10$^{-10}$, which are representative of helium-atmosphere
WDs. After the first iteration, if $M <$~0.406~\msun, we
switch to the evolutionary models of \citet{althaus01}, appropriate
for low-mass He-core WDs. In general, the \logg\ value obtained from
the inferred mass and radius $(g = GM/R^{2})$ will be different from
our initial guess of \logg\ =~8.0, and the fitting procedure is thus
repeated until an internal consistency in \logg\ is reached.

\begin{figure*}
\centering
\includegraphics[scale=0.4565,bb=20 87 567 687]{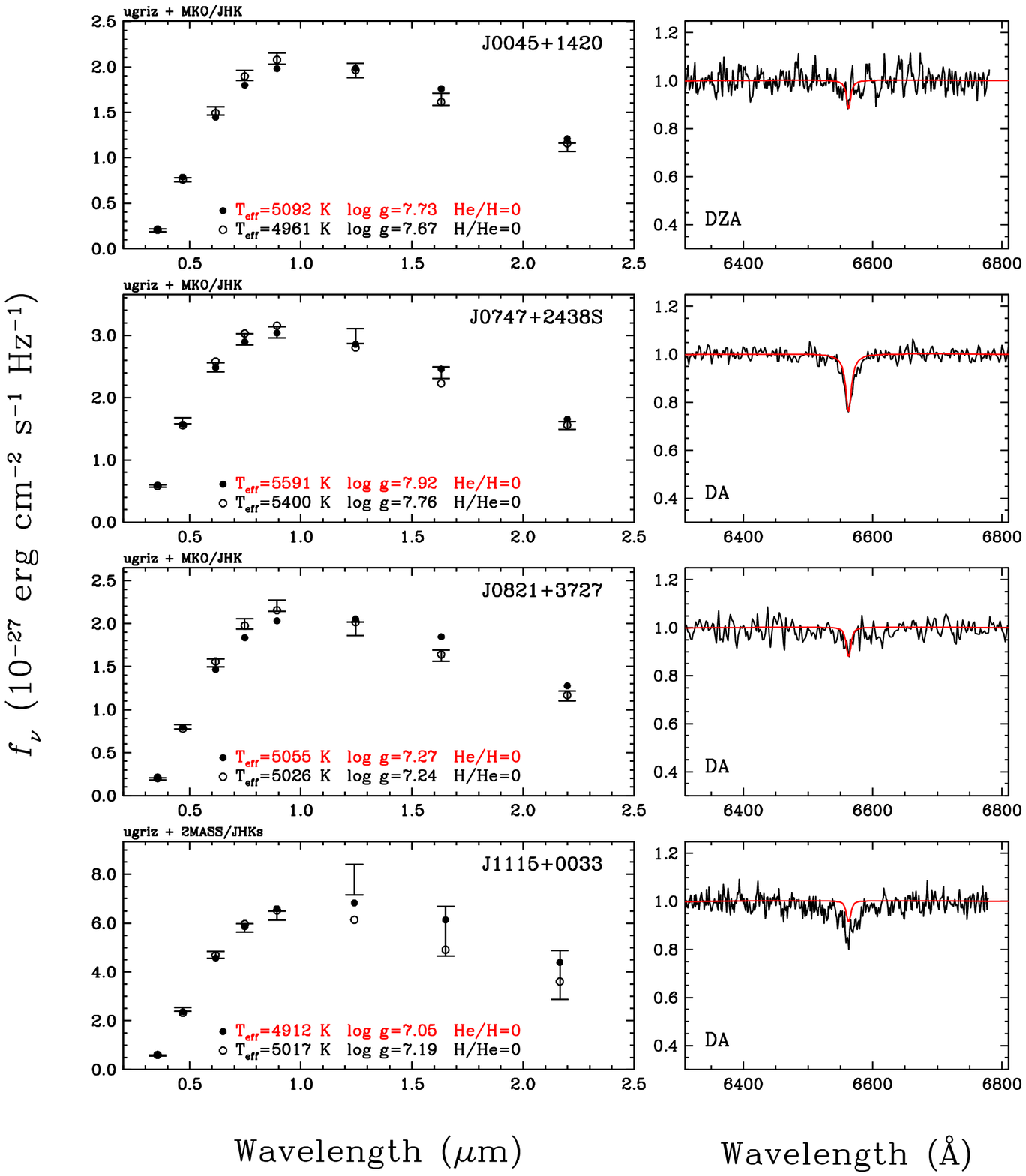}
\includegraphics[scale=0.4565,bb=20 87 567 687]{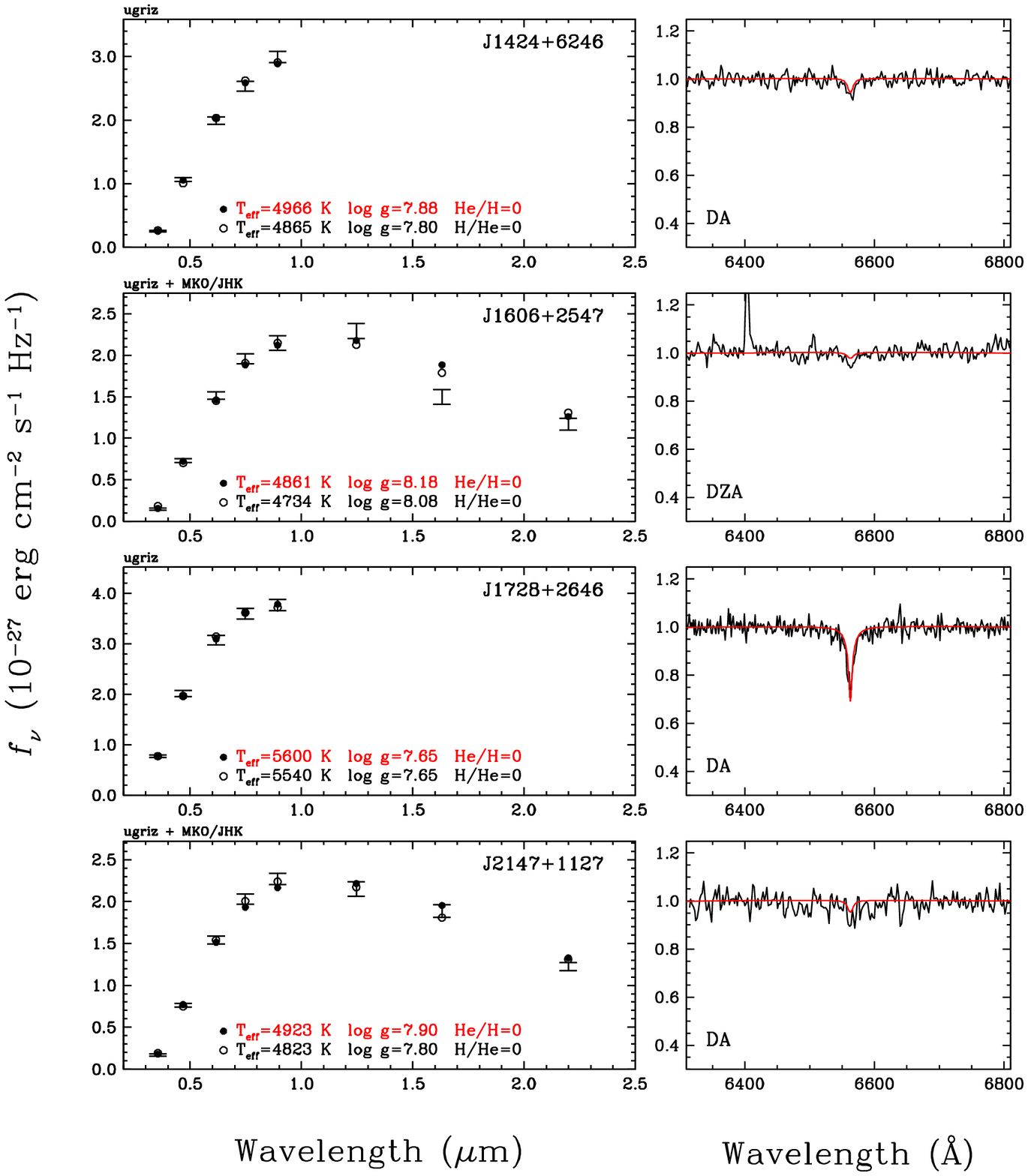}
\caption{Fits to the observed energy distributions with pure hydrogen
  models (filled circles) and with pure helium models (open circles)
  for the eight WDs that exhibit, or potentially exhibit, absorption
  at \halpha. Adopted atmospheric parameters are emphasized in
  red. Here and in the following figures, the photometric observations
  are represented by error bars while the filled and open circles
  represent the model fluxes for the pure H and pure He solution,
  respectively. In the right-hand panels we show the observed normalized
  spectra together with the synthetic line profiles calculated with
  the atmospheric parameters corresponding to the pure hydrogen
  solutions. \label{fg:DA}}
\end{figure*}

\subsection{Results}
Fig.~\ref{fg:color} presents the colour-magnitude diagram for our
parallax sample along with the evolutionary tracks for
0.3--0.9~\msun\ pure H, pure He, and 0.2~\msun\ mixed H/He atmosphere
models. Note that all the evolutionary tracks plotted in
Fig.~\ref{fg:color} represent the evolution of CO-core WDs. Two
other ultracool WDs with parallax measurements and SDSS photometry,
LHS 3250 and J1102 \citep{harris99,bergeron01,hall08,kilic12}, are
also included for comparison.

Interestingly, the majority of the targets in our sample fall above
the evolutionary tracks for 0.6~\msun\ WDs, indicating that they are
low-mass objects. Some of these WDs are even brighter than the
0.3~\msun\ WD sequence, implying masses as low as $\approx$~0.2~\msun.
A significant fraction of the stars in our sample are IR-faint WDs
that suffer from CIA from molecular hydrogen. The CIA affects the
redder optical bands and the infrared. Hence, most of these IR-faint
objects lie to the left of the pure H and pure He model
sequences. Note that our sample was selected to include as many
IR-faint WDs as possible. Therefore, these are overrepresented in
this figure. It is clear from this figure that the colour-magnitude
distribution of our sample is well matched by WD models with masses
$\approx$~0.2--0.9~\msun\ with a variety of compositions, including
pure H, pure He and mixed H/He atmospheres. Below we discuss the DA,
DC and ultracool WD samples separately.

\subsubsection{DA WDs}
Fig.~\ref{fg:DA} displays the best-fitting pure-hydrogen models to the
SEDs of the eight WDs classified as DA. Both the observed SEDs and the
\halpha\ line profiles are reproduced fairly well by our pure H
models. Given our parallax measurements, the best-fitting radii for these
eight targets range from 0.011 to 0.022~\rsun\ ($R>{\rm R}_{\earth}$),
indicating that they are relatively low-mass WDs. In fact, half of
these WDs have masses below 0.45~\msun, and therefore are likely
He-core WDs. The majority of low-mass WDs are in short-period ($P
\lesssim$ 1 d) binary systems \citep{marsh95,brown11}. Therefore,
J0045+1420, J0821+3727, J1115+0033, and J1728+2646 are likely
unresolved binary WDs. Table 4 provides WD cooling age estimates for
these DA WDs, as well as the rest of our parallax sample. For
$M<0.45$~\msun\ WDs, we provide cooling ages for both CO and He core
composition based on the evolutionary tracks of \citet{fontaine01} and
\citet{althaus01}, respectively. Regardless of the core composition,
these eight DA WDs have cooling ages of less than 8~Gyr.

It is necessary to note an important caveat regarding the four 
potential binaries listed above. If they are indeed unresolved binaries, 
then the WDs in these systems will be more massive than implied by our 
fits assuming a single star. Hence, their actual cooling ages will be 
larger for a given \Te. Our estimates for the cooling ages of these 
potential binaries should therefore be regarded, at best, as lower 
limits.

\subsubsection{DC WDs}
Fig.~\ref{fg:DC} shows our model fits to the SEDs of the 31 DC WDs
that are best explained by pure H or pure He atmosphere models. In all
cases, the optical spectra are featureless near the
\halpha\ region. Hence, the choice of a pure H or pure He composition
is based solely on the fits to the optical and infrared photometry. In
most cases, the atmospheric parameters from both the pure H and pure
He solution agree within the uncertainties.  Our model fits indicate
that all of these WDs have \Te\ $<$~5000~K. The ratio of the H to He
atmosphere WDs is 13/18. However, all DC WDs with temperatures below
\Te\ =~4530~K are best explained by H-rich atmosphere models
\citep[see also][]{kowalski06,giam12}.

Just like the DA sample discussed above, about half of the DCs in this
sample are low-mass objects.  The two coolest stars, J2118$-$0737 and
J2222+1221, have \Te~=~3920~$\pm$~60 and 4010~$\pm$~80~K, and 
$M$~=~0.31~$\pm$~0.09~\msun\ and 0.37~$\pm$~0.03~\msun, respectively.
Assuming He-cores, these temperatures correspond to cooling ages of
7.7 and 9.4 Gyr, respectively.  If these are short-period, unresolved
binary systems, then the companions would be fainter and more massive
WDs. Due to the unknown prior history of such binary systems and
without an estimate on their initial masses, their total ages,
including the main-sequence + WD cooling ages, cannot be reliably
calculated.

\begin{figure*}
\centering
\includegraphics[scale=0.58,angle=-90,bb=25 107 587 707]{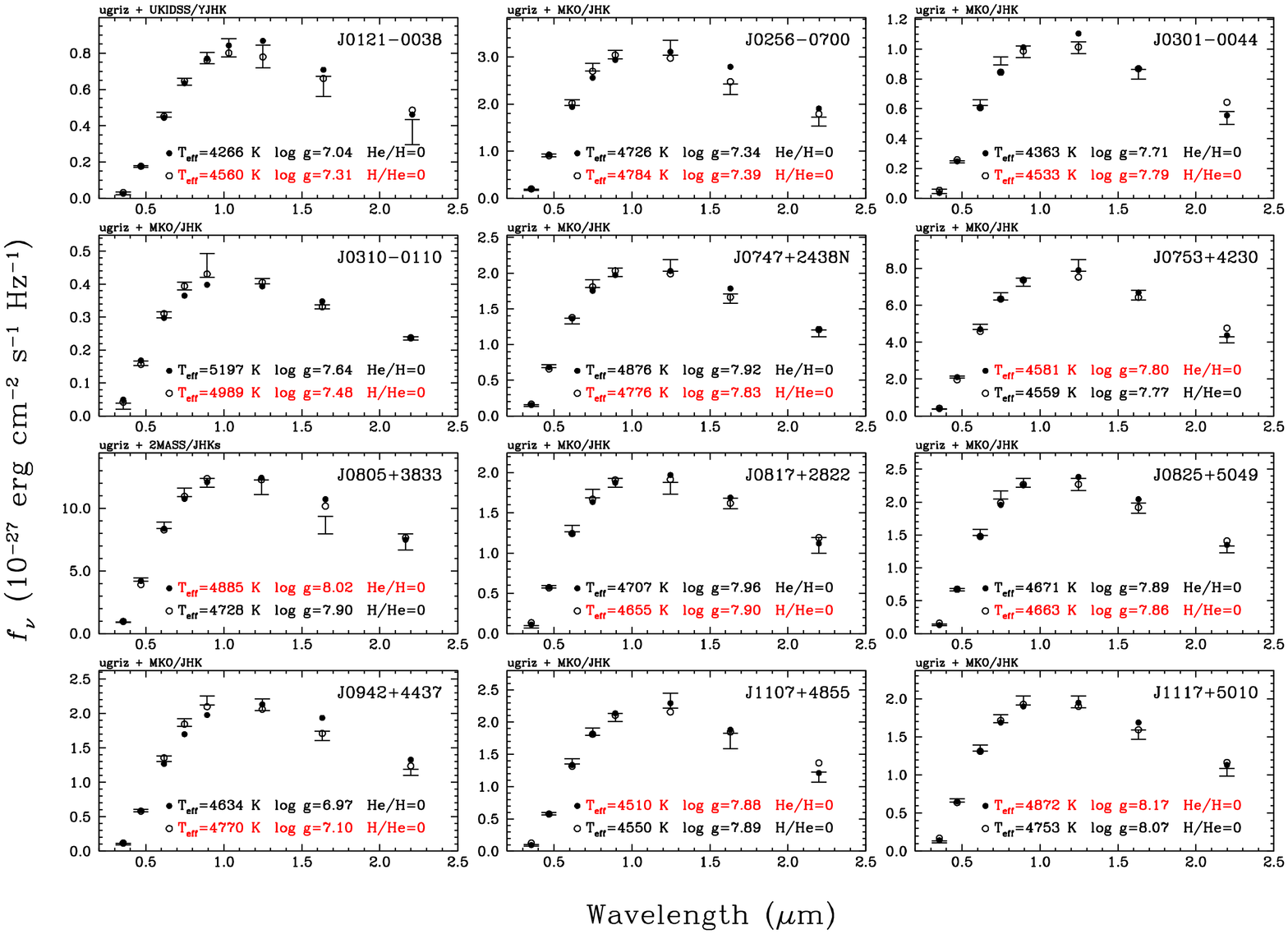}
\includegraphics[scale=0.58,angle=-90,bb=25 107 587 707]{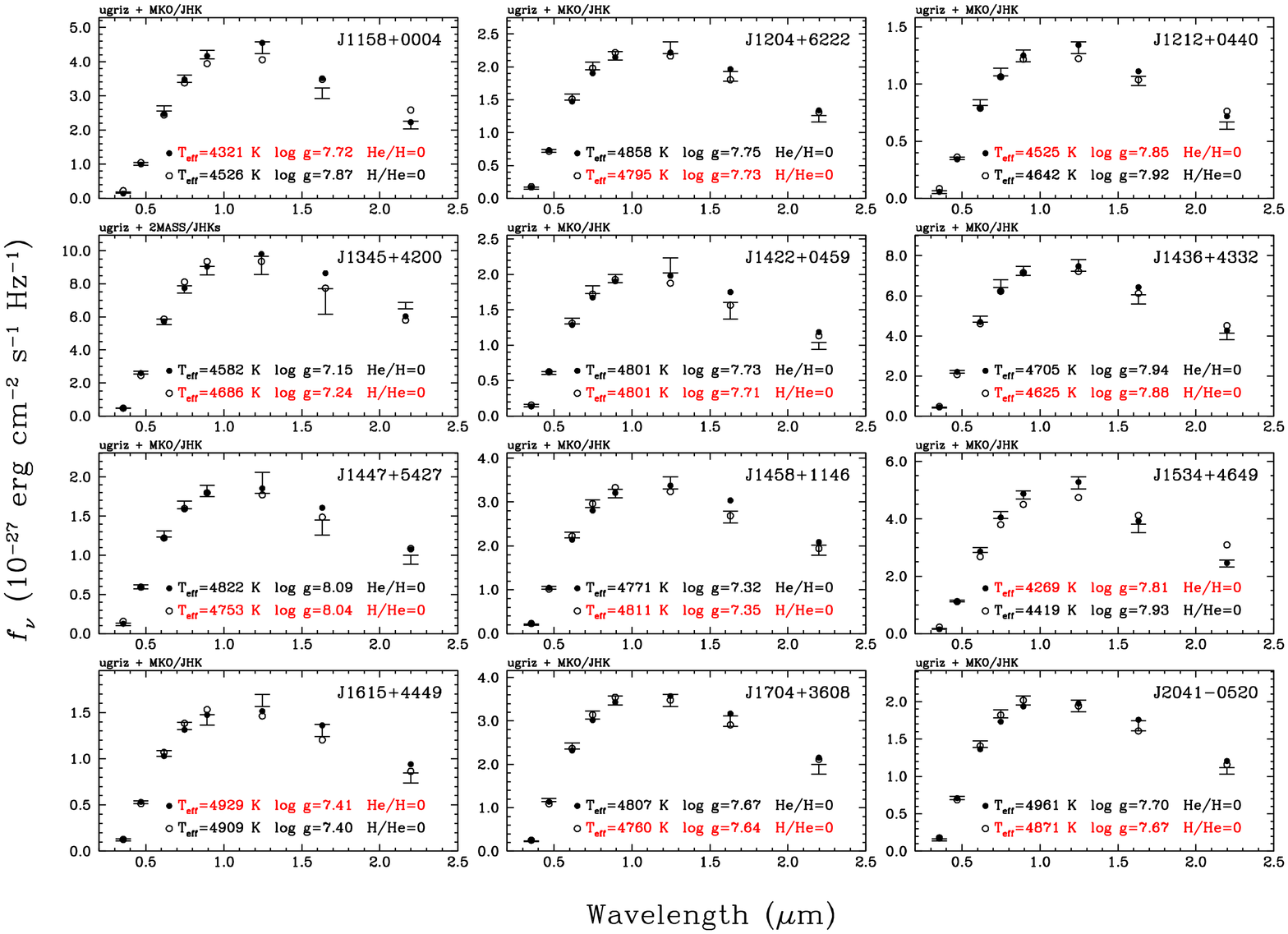}
\caption{Fits to the observed energy distributions with pure hydrogen
  models (filled circles) and with pure helium models (open circles)
  for the 31 DC WDs. All objects have featureless spectra near the
  \halpha\ region, and the SEDs are best explained with pure model
  atmospheres.  Adopted atmospheric parameters are emphasized in
  red. \label{fg:DC}}
\end{figure*}
\addtocounter{figure}{-1}
\begin{figure*}
\centering
\includegraphics[scale=0.58,angle=-90,bb=25 107 467 707]{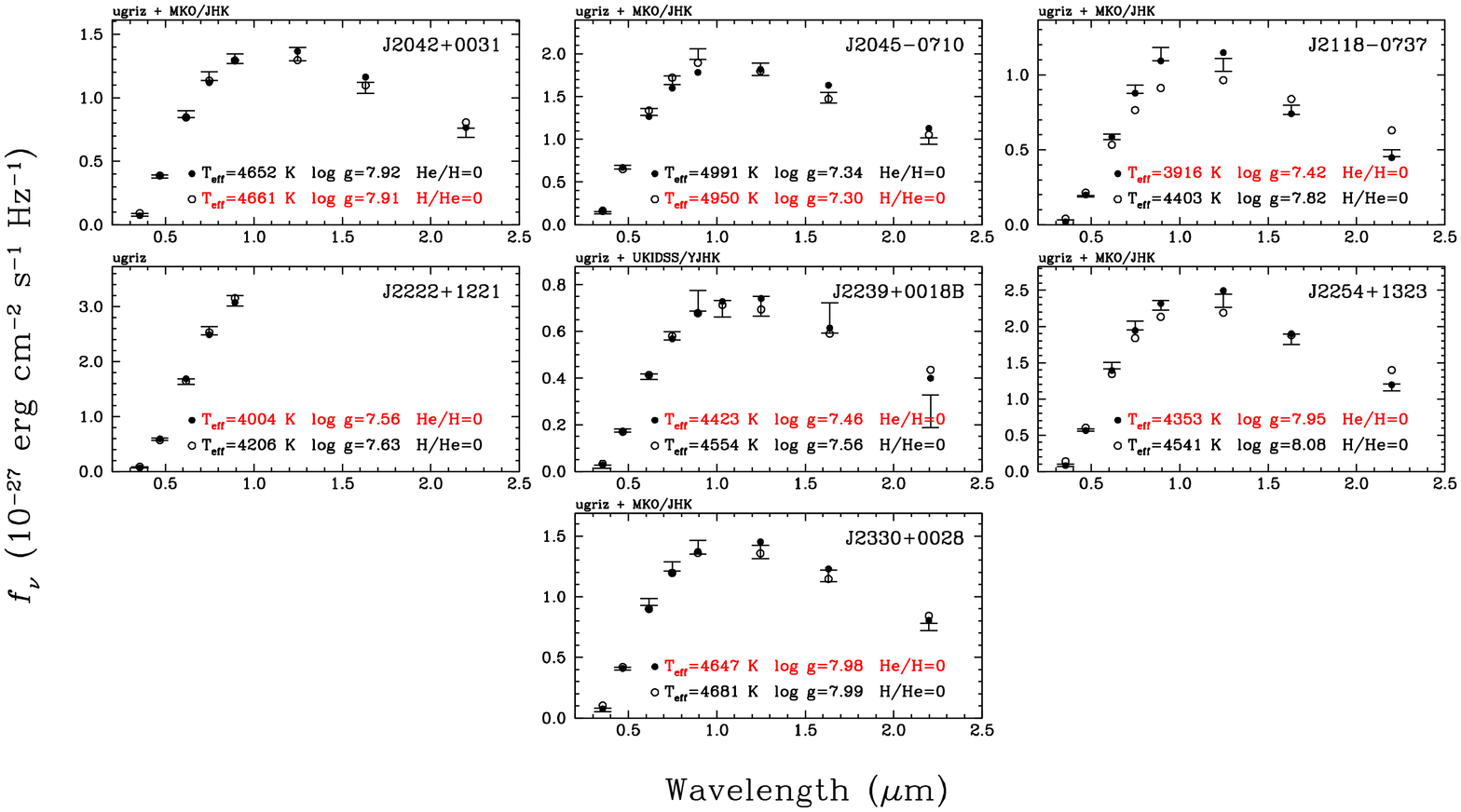}
\caption{ -- {\it continued}}
\end{figure*}

\subsubsection{DC WDs with Mixed H/He Atmospheres}
\citet{gates04}, \citet{harris08} and \citet{kilic10a} have
identified several IR-faint WDs that were originally thought to be
ultracool WDs with \Te\ $<$~4000~K. It turns out that some of these
IR-faint WDs are relatively warm. There are nine IR-faint, DC WDs in
our sample that are best-fitted with \Te\ $>$~4500~K mixed H/He
atmospheres models. The main opacity source in these mixed models is
the H$_{2}$--He CIA in the infrared. Since cool He-rich WDs have lower
opacities and higher atmospheric pressures, the CIA becomes effective
at higher temperatures \citep[\Te\ $>$~4000~K,][]{bergeron02}.

\begin{figure*}
\centering
\includegraphics[scale=0.615,angle=-90,bb=20 87 587 707]{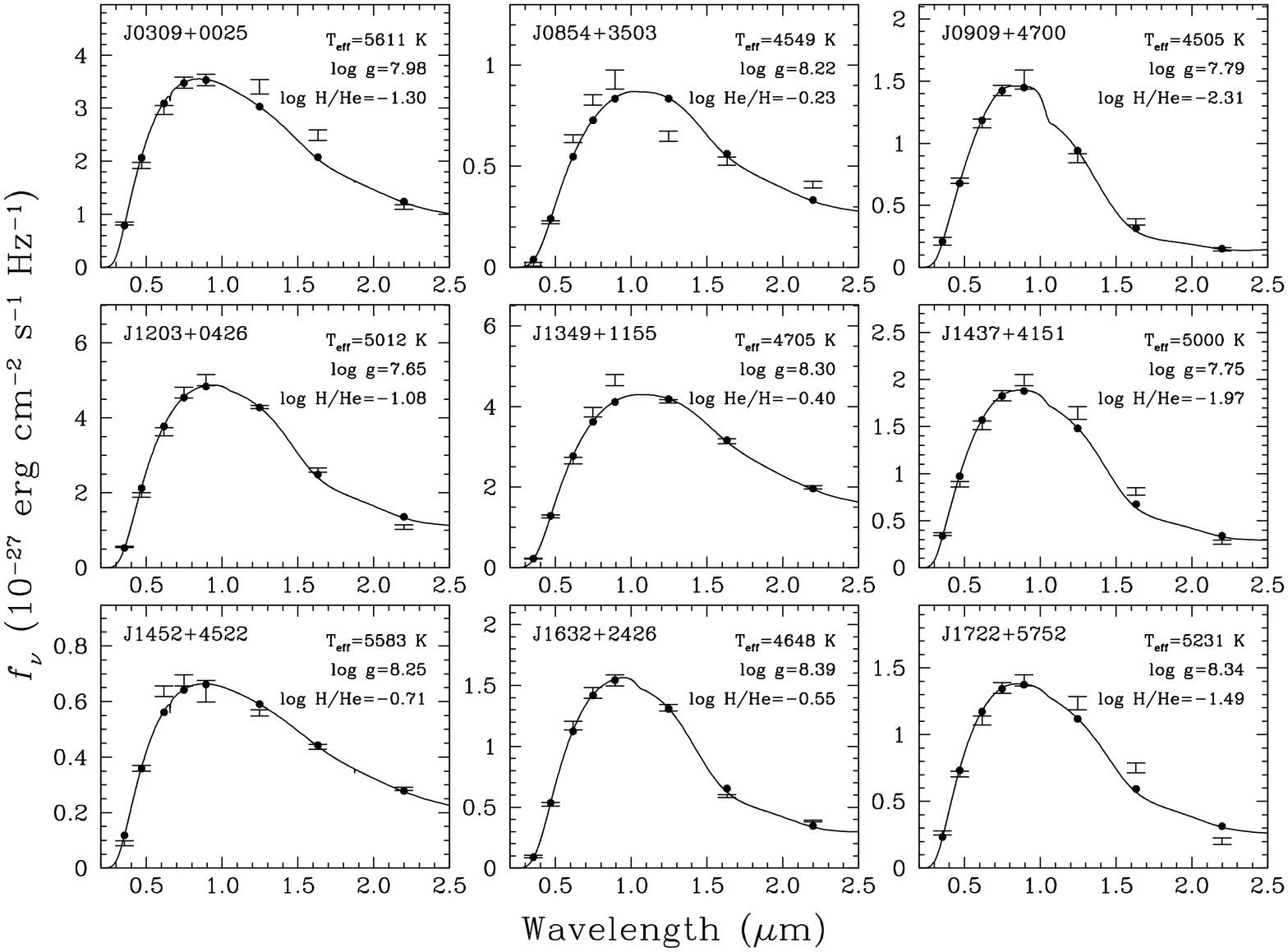}
\caption{Fits to the SEDs of the nine IR-faint, DC WDs in our sample,
  excluding the ultracool WDs. All objects have featureless spectra
  near the \halpha\ region, and the SEDs are best explained with mixed
  model atmospheres. Note that the measured abundances are quoted
  relative to the dominant atmospheric constituent.\label{fg:mix}}
\end{figure*}

Fig.~\ref{fg:mix} shows the SEDs for these nine DC WDs with mixed
composition. The mixed models with $\log{({\rm He/H})} = -0.4$ to 2.3
fit the observed SEDs (over the 0.3--2.2 $\mu$m region) fairly
well. The best-fitting parameters for some of these stars are markedly
different than the parameters presented in \citet{kilic10a}. However,
the analysis presented in this paper is superior to earlier work since
we now include all available photometry in our analysis (including the
$u$-band data) and we also have parallax measurements available.
J1632+2426 is the most-massive and the oldest WD (in terms of the WD
cooling age) in this sample, with a mass of 0.82~$\pm$~0.04~\msun\ and
a cooling age of 7.7~Gyr.

\begin{figure*}
\centering
\includegraphics[scale=0.62,angle=-90,bb=100 87 507 707]{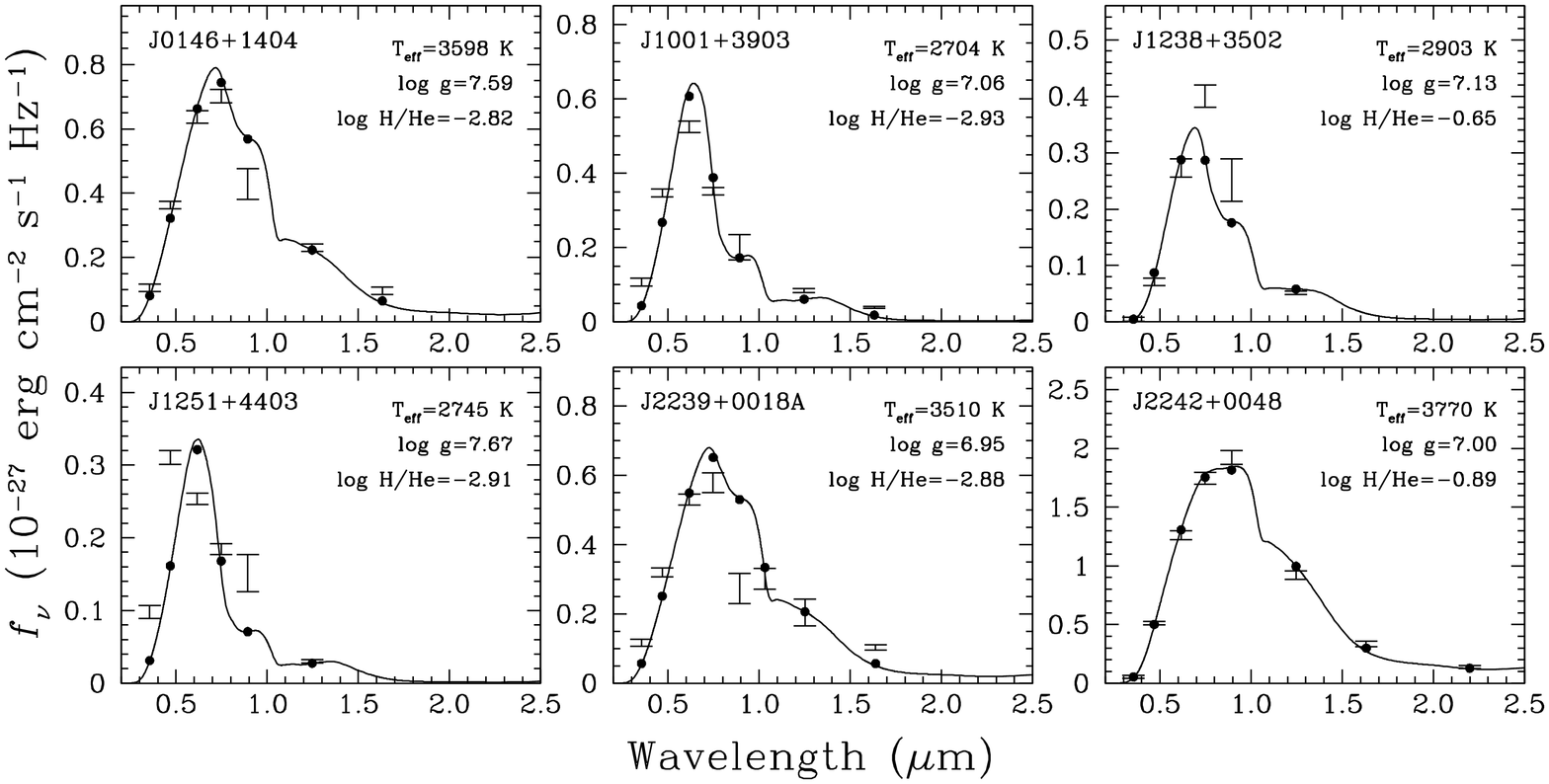}
\caption{Fits to the SEDs of the six ultracool DC WDs in our
  sample. All objects have featureless spectra near the
  \halpha\ region, and the SEDs are best explained with mixed model
  atmospheres. Note that the measured abundances are quoted relative
  to the dominant atmospheric constituent.\label{fg:UC}}
\end{figure*}

\subsubsection{Ultracool WDs}

We originally selected 12 ultracool WD candidates for follow-up
parallax observations: J0854+3503 and J1001+3903 from \citet{gates04};
J0121$-$0038, J0301$-$0044, J2239+0018 and J2242+0048 from
\citet{vidrih07}; J0146+1404, J0310$-$0110, J1238+3502, J1251+4403,
J1452+4522 and J1632+2426 from \citet{harris08}. Our detailed model
atmosphere analysis using parallax data shows that only half of these
stars are actually ultracool WDs with \Te\ $<$~4000~K. The rest of the
ultracool candidates are best explained by pure H/He or mixed
atmosphere models with \Te\ $>$~4000~K.

Fig.~\ref{fg:UC} shows the SEDs and our model fits to the six
ultracool WDs in our sample. The best-fitting parameters are given in each
panel and at the end of Table 4. Note that prior to this work, there
were only three ultracool WDs with parallax observations
available. Hence, the ultracool WD sample presented here is a
significant addition to this sample. The six ultracool WDs presented
here are best explained by mixed H/He atmospheres with
\Te\ =~2710--3760~K and $\log{({\rm
    He/H})}$~=~0.65--2.96. Interestingly, all six of these ultracool
WDs are too bright for average mass WDs. Instead, the observed
parallaxes require relatively large radii ($R$~=~0.015--0.023~\rsun)
and low masses ($M$~=~0.17--0.39~\msun). Assuming He-cores, the WD
cooling ages range from 4.5 to 9.7~Gyr. They are located within
63--110~pc of the Sun and they display tangential velocities of
40--140~\kms. Hence, these ultracool WDs likely belong to the 
Galactic disc.

\setlength{\tabcolsep}{6pt}
\setlength{\extrarowheight}{1.5pt}
\begin{table*}
\scriptsize
\centering
\begin{minipage}{.85\textwidth}
\caption{Properties of cool WDs.}
\begin{tabular}{@{}lr@{ $\pm$ }rr@{ $\pm$ }lr@{ $\pm$ }lr@{ $\pm$ }lcr@{ $\pm$ }lr@{}lr@{}l}
\hline
\noalign{\smallskip}
SDSS     & \multicolumn{2}{c}{\Te} & \multicolumn{2}{c}{\logg}        & \multicolumn{2}{c}{$M$}     & \multicolumn{2}{c}{$R$}      & Comp.         & \multicolumn{2}{c}{$M_{\rm g}$} & \multicolumn{2}{c}{$\tau_{\rm cool, CO}$} & \multicolumn{2}{c}{$\tau_{\rm cool, He}$} \\
         & \multicolumn{2}{c}{(K)} & \multicolumn{2}{c}{(cm s$^{-2}$)} & \multicolumn{2}{c}{(\msun)} &  \multicolumn{2}{c}{(\rsun)} & ($\log$ He/H) & \multicolumn{2}{c}{(mag)} & \multicolumn{2}{c}{(Gyr)}             & \multicolumn{2}{c}{(Gyr)} \\
\noalign{\smallskip}
\hline
\multicolumn{16}{c}{DA}\\
\hline
\noalign{\smallskip}
J0045+1420   & 5090 &  60 & 7.73 & 0.15 & 0.43 & 0.07 & 0.015 & 0.001 &    H  & 15.10 & 0.18 & 2.7 & $^{+0.9}_{-0.6}$ & \multicolumn{2}{c}{--} \\
J0747+2438S  & 5590 &  40 & 7.92 & 0.09 & 0.54 & 0.05 & 0.013 & 0.001 &    H  & 14.67 & 0.12 & 2.5 & $^{+0.5}_{-0.3}$ & \multicolumn{2}{c}{--} \\
J0821+3727   & 5050 &  50 & 7.27 & 0.50 & 0.27 & 0.15 & 0.020 & 0.006 &    H  & 14.47 & 0.59 & 1.6 & $^{+1.0}_{-1.3}$ & 3.1 & $^{+2.4}_{-1.1}$ \\
J1115+0033   & 4910 &  40 & 7.05 & 0.27 & 0.21 & 0.07 & 0.022 & 0.003 &    H  & 14.38 & 0.28 & 1.4 & $^{+0.4}_{-0.7}$ & 2.9 & $^{+0.6}_{-0.6}$ \\
J1424+6246   & 4970 &  60 & 7.88 & 0.16 & 0.51 & 0.09 & 0.014 & 0.001 &    H  & 15.42 & 0.20 & 4.4 & $^{+1.7}_{-1.4}$ & \multicolumn{2}{c}{--} \\
J1606+2547   & 4860 &  60 & 8.18 & 0.12 & 0.69 & 0.08 & 0.011 & 0.001 &    H  & 15.97 & 0.17 & 7.9 & $^{+0.6}_{-1.1}$ & \multicolumn{2}{c}{--} \\
J1728+2646   & 5600 &  50 & 7.65 & 0.17 & 0.42 & 0.07 & 0.016 & 0.002 &    H  & 14.29 & 0.24 & 1.6 & $^{+0.4}_{-0.2}$ & 3.5 & $^{+0.8}_{-0.6}$ \\
J2147+1127   & 4920 &  60 & 7.90 & 0.18 & 0.52 & 0.10 & 0.013 & 0.002 &    H  & 15.53 & 0.24 & 4.9 & $^{+1.8}_{-1.8}$ & \multicolumn{2}{c}{--} \\
\noalign{\smallskip}
\hline
\multicolumn{16}{c}{DC}\\
\hline
\noalign{\smallskip}
J0121$-$0038 & 4560 &  50 & 7.31 & 0.39 & 0.28 & 0.13 & 0.019 & 0.004 &   He  & 15.43 & 0.46 & 2.2 & $^{+1.1}_{-1.2}$ & 4.6 & $^{+3.2}_{-1.6}$ \\
J0256$-$0700 & 4780 &  40 & 7.39 & 0.15 & 0.30 & 0.05 & 0.018 & 0.002 &   He  & 15.07 & 0.20 & 2.2 & $^{+0.3}_{-0.3}$ & 4.6 & $^{+1.4}_{-0.9}$ \\
J0301$-$0044 & 4530 &  50 & 7.79 & 0.16 & 0.45 & 0.08 & 0.014 & 0.001 &   He  & 16.08 & 0.19 & 4.8 & $^{+1.6}_{-1.2}$ & \multicolumn{2}{c}{--} \\
J0310$-$0110 & 4990 &  50 & 7.48 & 0.47 & 0.34 & 0.17 & 0.018 & 0.005 &   He  & 14.82 & 0.59 & 2.1 & $^{+2.1}_{-1.1}$ & 4.7 & $^{+3.2}_{-2.0}$ \\
J0747+2438N  & 4780 &  40 & 7.83 & 0.10 & 0.47 & 0.05 & 0.014 & 0.001 &   He  & 15.59 & 0.12 & 4.5 & $^{+1.1}_{-0.9}$ & \multicolumn{2}{c}{--} \\
J0753+4230   & 4580 &  40 & 7.80 & 0.05 & 0.46 & 0.03 & 0.014 & 0.001 &    H  & 15.89 & 0.06 & 5.1 & $^{+0.6}_{-0.6}$ & \multicolumn{2}{c}{--} \\
J0805+3833   & 4890 &  40 & 8.02 & 0.03 & 0.59 & 0.02 & 0.012 & 0.001 &    H  & 15.70 & 0.05 & 6.4 & $^{+0.3}_{-0.4}$ & \multicolumn{2}{c}{--} \\
J0817+2822   & 4660 &  50 & 7.90 & 0.13 & 0.51 & 0.08 & 0.013 & 0.001 &   He  & 15.91 & 0.17 & 5.6 & $^{+1.5}_{-1.3}$ & \multicolumn{2}{c}{--} \\
J0825+5049   & 4660 &  40 & 7.86 & 0.11 & 0.49 & 0.06 & 0.014 & 0.001 &   He  & 15.89 & 0.14 & 5.2 & $^{+1.3}_{-1.0}$ & \multicolumn{2}{c}{--} \\
J0942+4437   & 4770 &  40 & 7.10 & 0.21 & 0.22 & 0.06 & 0.022 & 0.002 &   He  & 14.77 & 0.23 & 1.6 & $^{+0.3}_{-0.6}$ & 3.3 & $^{+0.6}_{-0.5}$ \\
J1107+4855   & 4510 &  70 & 7.88 & 0.14 & 0.50 & 0.08 & 0.014 & 0.001 &    H  & 16.08 & 0.18 & 6.3 & $^{+1.5}_{-1.6}$ & \multicolumn{2}{c}{--} \\
J1117+5010   & 4870 &  60 & 8.17 & 0.12 & 0.69 & 0.08 & 0.011 & 0.001 &    H  & 15.93 & 0.18 & 7.8 & $^{+0.8}_{-1.1}$ & \multicolumn{2}{c}{--} \\
J1158+0004   & 4320 &  70 & 7.72 & 0.11 & 0.42 & 0.05 & 0.015 & 0.001 &    H  & 16.16 & 0.13 & 5.2 & $^{+1.1}_{-1.0}$ & \multicolumn{2}{c}{--} \\
J1204+6222   & 4800 &  40 & 7.73 & 0.23 & 0.44 & 0.11 & 0.015 & 0.003 &   He  & 15.43 & 0.31 & 7.2 & $^{+4.3}_{-2.0}$ & \multicolumn{2}{c}{--} \\
J1212+0440   & 4530 &  80 & 7.85 & 0.29 & 0.49 & 0.16 & 0.014 & 0.003 &    H  & 16.01 & 0.37 & 5.9 & $^{+2.5}_{-2.8}$ & \multicolumn{2}{c}{--} \\
J1345+4200   & 4690 &  20 & 7.25 & 0.06 & 0.26 & 0.02 & 0.020 & 0.001 &   He  & 15.02 & 0.08 & 2.0 & $^{+0.1}_{-0.1}$ & 4.0 & $^{+0.3}_{-0.3}$ \\
J1422+0459   & 4800 &  50 & 7.71 & 0.24 & 0.41 & 0.11 & 0.015 & 0.002 &   He  & 15.51 & 0.29 & 3.5 & $^{+2.2}_{-1.0}$ & \multicolumn{2}{c}{--} \\
J1436+4332   & 4630 &  30 & 7.88 & 0.05 & 0.50 & 0.03 & 0.013 & 0.001 &   He  & 15.88 & 0.07 & 5.4 & $^{+0.6}_{-0.6}$ & \multicolumn{2}{c}{--} \\
J1447+5427   & 4750 &  60 & 8.04 & 0.30 & 0.59 & 0.18 & 0.012 & 0.003 &   He  & 15.92 & 0.40 & 6.9 & $^{+0.5}_{-3.5}$ & \multicolumn{2}{c}{--} \\
J1458+1146   & 4810 &  40 & 7.35 & 0.15 & 0.29 & 0.05 & 0.019 & 0.002 &   He  & 14.96 & 0.20 & 2.1 & $^{+0.3}_{-0.3}$ & 4.2 & $^{+1.3}_{-0.8}$ \\
J1534+4649   & 4270 &  60 & 7.82 & 0.09 & 0.47 & 0.05 & 0.014 & 0.001 &    H  & 16.35 & 0.12 & 6.4 & $^{+1.0}_{-1.0}$ & \multicolumn{2}{c}{--} \\
J1615+4449   & 4930 &  60 & 7.41 & 0.57 & 0.31 & 0.18 & 0.018 & 0.007 &    H  & 14.84 & 0.74 & 2.0 & $^{+1.9}_{-1.1}$ & 4.3 & $^{+3.3}_{-2.3}$ \\
J1704+3608   & 4760 &  40 & 7.64 & 0.13 & 0.40 & 0.06 & 0.016 & 0.001 &   He  & 15.31 & 0.18 & 2.9 & $^{+0.9}_{-0.4}$ & 6.8 & $^{+1.2}_{-1.1}$ \\
J2041$-$0520 & 4870 &  40 & 7.67 & 0.17 & 0.42 & 0.07 & 0.016 & 0.002 &   He  & 15.20 & 0.23 & 2.8 & $^{+1.1}_{-0.6}$ & 6.4 & $^{+1.5}_{-1.3}$ \\
J2042+0031   & 4660 &  50 & 7.91 & 0.15 & 0.52 & 0.09 & 0.013 & 0.001 &   He  & 15.95 & 0.19 & 5.7 & $^{+1.5}_{-1.6}$ & \multicolumn{2}{c}{--} \\
J2045$-$0710 & 4950 &  40 & 7.30 & 0.21 & 0.28 & 0.06 & 0.019 & 0.003 &   He  & 14.66 & 0.26 & 1.8 & $^{+0.4}_{-0.4}$ & 3.5 & $^{+1.4}_{-0.6}$ \\
J2118$-$0737 & 3920 &  60 & 7.42 & 0.25 & 0.31 & 0.09 & 0.018 & 0.003 &    H  & 16.41 & 0.32 & 3.4 & $^{+1.8}_{-0.8}$ & 7.7 & $^{+3.6}_{-2.3}$ \\
J2222+1221   & 4010 &  80 & 7.56 & 0.08 & 0.37 & 0.03 & 0.017 & 0.001 &    H  & 16.42 & 0.11 & 4.2 & $^{+0.9}_{-0.6}$ & 9.4 & $^{+1.3}_{-1.3}$ \\
J2239+0018B  & 4420 &  90 & 7.46 & 0.55 & 0.33 & 0.19 & 0.018 & 0.006 &    H  & 15.64 & 0.66 & 2.9 & $^{+3.9}_{-2.2}$ & 6.3 & $^{+7.2}_{-3.4}$ \\
J2254+1323   & 4350 &  70 & 7.95 & 0.12 & 0.54 & 0.07 & 0.013 & 0.001 &    H  & 16.39 & 0.16 & 7.5 & $^{+1.1}_{-1.3}$ & \multicolumn{2}{c}{--} \\
J2330+0028   & 4650 &  80 & 7.98 & 0.22 & 0.57 & 0.13 & 0.013 & 0.002 &    H  & 16.03 & 0.29 & 6.9 & $^{+1.7}_{-2.7}$ & \multicolumn{2}{c}{--} \\
\noalign{\smallskip}
\hline
\multicolumn{16}{c}{Mixed}\\
\hline
\noalign{\smallskip}
J0309+0025   & 5610 &  70 & 7.98 & 0.13 & 0.56 & 0.08 & 0.013 & 0.001 &   1.30& 14.74 & 0.18 & 3.4 & $^{+1.1}_{-0.8}$ & \multicolumn{2}{c}{--} \\
J0854+3503   & 4550 &  80 & 8.22 & 0.15 & 0.71 & 0.10 & 0.011 & 0.001 &$-$0.23& 16.50 & 0.22 & 7.6 & $^{+0.2}_{-0.1}$ & \multicolumn{2}{c}{--} \\
J0909+4700   & 4510 & 110 & 7.79 & 0.19 & 0.45 & 0.10 & 0.014 & 0.002 &  2.31 & 15.36 & 0.23 & 4.8 & $^{+2.0}_{-1.4}$ & \multicolumn{2}{c}{--} \\
J1203+0426   & 5010 &  40 & 7.65 & 0.16 & 0.41 & 0.07 & 0.016 & 0.002 &  1.08 & 14.91 & 0.22 & 5.6 & $^{+1.0}_{-1.0}$ & \multicolumn{2}{c}{--} \\
J1349+1155   & 4710 &  40 & 8.30 & 0.06 & 0.76 & 0.04 & 0.010 & 0.001 &$-$0.40& 16.37 & 0.10 & 7.5 & $^{+0.1}_{-0.1}$ & \multicolumn{2}{c}{--} \\
J1437+4151   & 5000 &  70 & 7.75 & 0.24 & 0.45 & 0.11 & 0.015 & 0.003 &  1.97 & 14.93 & 0.32 & 6.2 & $^{+4.9}_{-1.6}$ & \multicolumn{2}{c}{--} \\
J1452+4522   & 5580 &  70 & 8.25 & 0.20 & 0.73 & 0.13 & 0.011 & 0.002 &  0.71 & 15.12 & 0.30 & 5.5 & $^{+0.6}_{-1.5}$ & \multicolumn{2}{c}{--} \\
J1632+2426   & 4650 &  70 & 8.39 & 0.06 & 0.82 & 0.04 & 0.010 & 0.001 &  0.55 & 16.38 & 0.10 & 7.7 & $^{+0.1}_{-0.1}$ & \multicolumn{2}{c}{--} \\
J1722+5752   & 5230 &  80 & 8.34 & 0.14 & 0.79 & 0.09 & 0.010 & 0.001 &  1.49 & 15.54 & 0.21 & 6.7 & $^{+0.1}_{-0.5}$ & \multicolumn{2}{c}{--} \\
\noalign{\smallskip}
\hline
\multicolumn{16}{c}{Ultracool}\\
\hline
\noalign{\smallskip}
J0146+1404   & 3600 & 140 & 7.59 & 0.22 & 0.38 & 0.09 & 0.016 & 0.003 &  2.82 & 15.23 & 0.29 & 3.7 & $^{+1.7}_{-1.0}$ & 9.2 & $^{+3.1}_{-3.2}$ \\
J1001+3903   & 2710 & 150 & 7.06 & 0.24 & 0.20 & 0.09 & 0.022 & 0.004 &  2.93 & 15.35 & 0.39 & 2.9 & $^{+2.3}_{-0.7}$ & 7.6 & $^{+4.2}_{-1.7}$ \\
J1238+3502   & 2900 & 210 & 7.13 & 0.42 & 0.22 & 0.11 & 0.021 & 0.004 &  0.65 & 16.77 & 0.44 & 3.6 & $^{+3.2}_{-0.1}$ & 9.7 & $^{+6.8}_{-5.0}$ \\
J1251+4403   & 2750 & 180 & 7.67 & 0.69 & 0.39 & 0.28 & 0.015 & 0.006 &  2.91 & 16.17 & 0.73 & 9.1 & $^{+1.4}_{-4.9}$ & 9.1 & $^{+1.3}_{-5.0}$ \\
J2239+0018A  & 3510 & 220 & 6.95 & 0.34 & 0.20 & 0.09 & 0.025 & 0.008 &  2.88 & 14.99 & 0.66 & 2.0 & $^{+1.8}_{-0.9}$ & 5.2 & $^{+3.7}_{-1.5}$ \\
J2242+0048   & 3770 &  90 & 7.00 & 0.13 & 0.20 & 0.03 & 0.023 & 0.003 &  0.89 & 15.41 & 0.25 & 1.9 & $^{+0.8}_{-1.2}$ & 4.8 & $^{+1.1}_{-0.6}$ \\
\noalign{\smallskip}
\hline
\label{tab:tg}
\end{tabular}
\end{minipage}
\end{table*}

\section{Discussion}

\subsection{Nearby WDs}
The local WD population is complete to within 13~pc, and there remains
a significant number of WDs to be discovered in the solar neighbourhood
\citep{holberg08,giam12}. Through our parallax observations, here we
have uncovered WDs with distances ranging from 21 to
$\approx$100~pc.  \citet{sion14} present 224 WDs within 25~pc of the
Sun. With a distance of 21~$\pm$~1~pc, J0805+3833 (WD~0802+387) is a
new addition to this sample. There are also four other WDs,
J0753+4230, J1349+1155, J1436+4332 and J1534+4649, with distances
$\leq$~30~pc. Since parallax observations on individual targets is
time consuming, significant progress on creating a complete sample of
WDs in the solar neighbourhood has to wait until astrometric data from
large scale surveys such as $GAIA$ \citep{perryman01} and the Large
Synoptic Survey Telescope \citep[LSST;][]{ivezic08} become available.

\begin{figure}
\centering
\includegraphics[scale=0.55,angle=-90,bb=102 136 496 684]{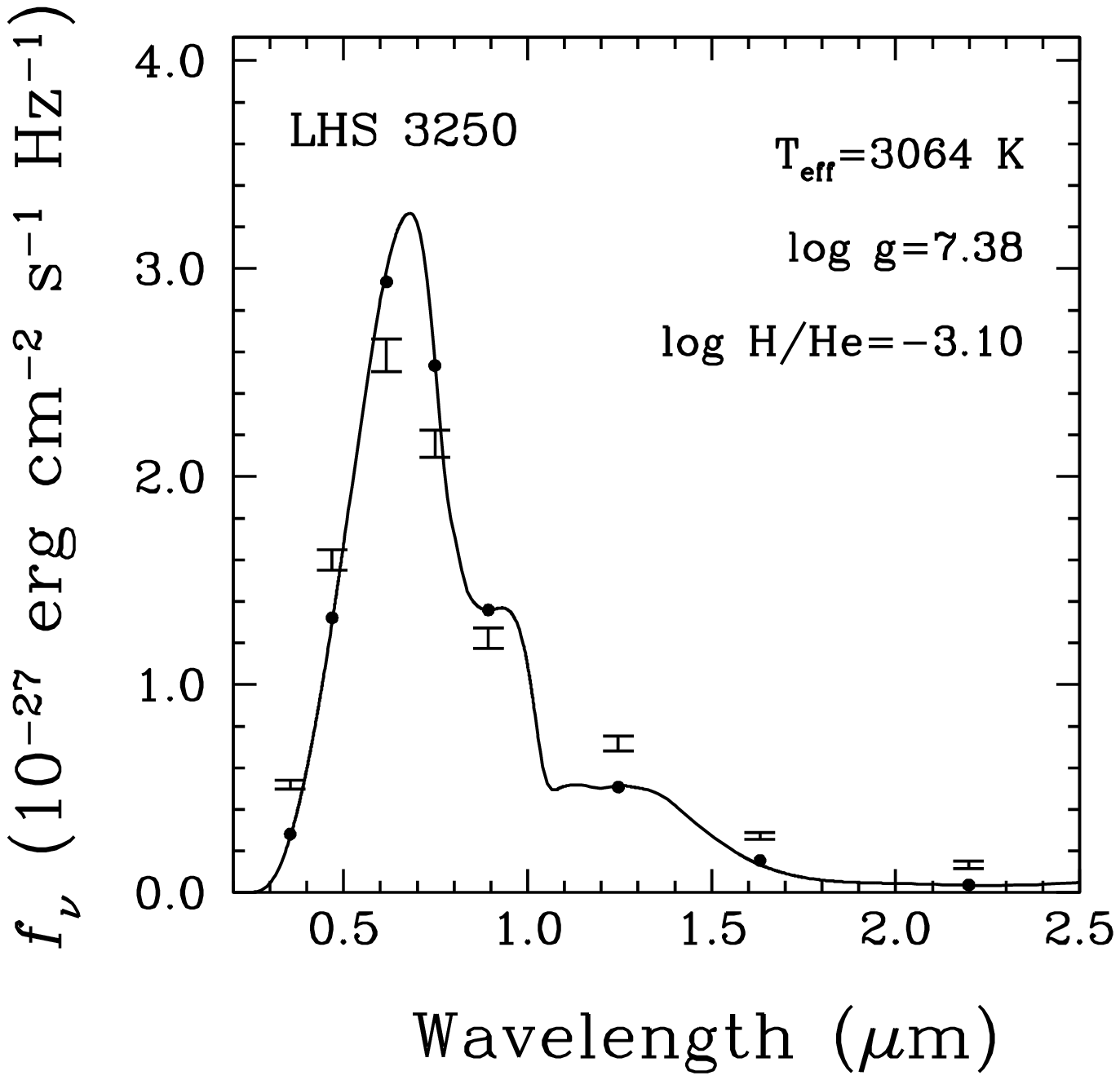}
\caption{Similar to Fig.~\ref{fg:UC} but for the ultracool WD
  LHS~3250.
\label{fg:LHS}}
\end{figure}

\begin{figure}
\centering
\includegraphics[scale=0.55,angle=-90,bb=102 136 496 684]{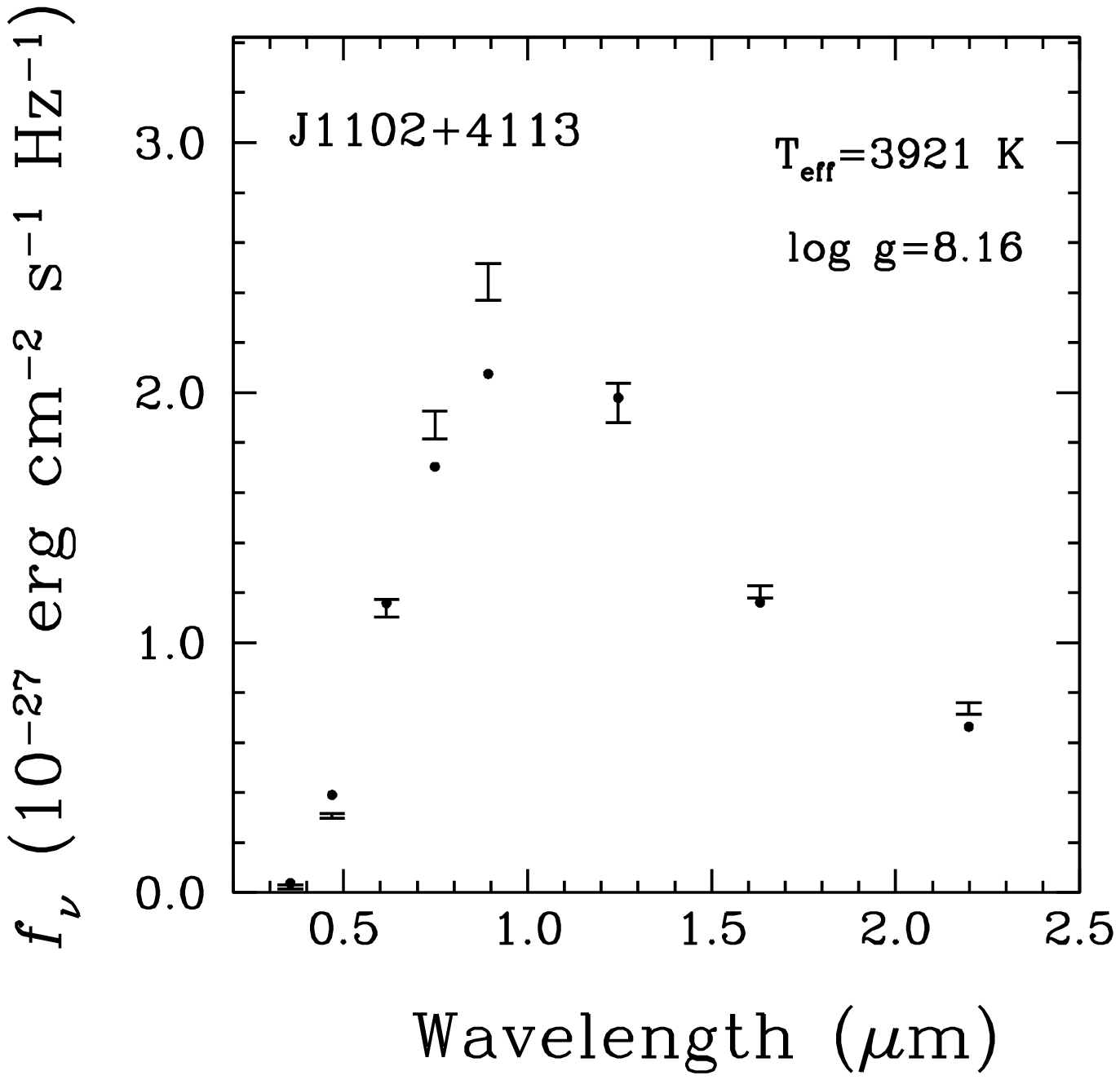}
\caption{Similar to Fig.~\ref{fg:DC} but for the ultracool WD
  J1102+4113 where we show the pure H solution only.
\label{fg:J1102}}
\end{figure}

\subsection{The Nature of Ultracool WDs}
The colour-magnitude diagram presented in Fig.~\ref{fg:color} shows
that the faintest WDs in our sample have $M_r \leq 16.4$ mag. This
magnitude limit corresponds to WD cooling ages of 11 Gyr for
0.6~\msun\ CO-core and pure H atmosphere WDs. There are six ultracool
WDs in our sample with best-fitting temperatures of $<$~4000~K.
Constraining the nature, including the total ages of these stars, has
been problematic.

Previously, \citet{oppen01}, \citet{bergeron01}, and \citet{kilic12}
presented detailed model atmosphere analysis of three ultracool WDs
with parallax measurements. WD~0346+246 is a 3650~K, \logg\ =~8.3
mixed atmosphere WD with $\log{{(\rm He/H)}}$~=~0.43, whereas J1102 is
best explained by a pure H atmosphere model with \Te\ =~3830~K and
\logg\ =~8.08 \citep{kilic12}. These two ultracool WDs have masses of
0.77 and 0.62~\msun, respectively. Their total main-sequence + WD
cooling ages and their kinematic properties indicate Galactic halo
membership. LHS~3250 stands out in this sample. \citet{bergeron02}
find a best-fitting solution of \Te\ =~3042~K, \logg\ =~7.27, and
$\log{({\rm H/He})}=-2.7$.  Fig.~\ref{fg:LHS} shows our model fits
to the LHS 3250 SED. Including the red wing of the Ly$\alpha$ opacity
\citep{kowalski06}, we now derive slightly different parameters of
\Te\ =~3064~K, \logg\ =~7.38, and $\log{({\rm H/He})}=-3.1$. Even
though the models predict deep absorption features around 0.8 and
1.1~$\mu$m, these features have never been observed in the actual
spectra of cool and ultracool WDs. Clearly, the models have problems
(likely due to problems with CIA calculations).  Nevertheless,
LHS~3250 is too bright compared to the pure H atmosphere models for
ultracool WDs (see Fig.~\ref{fg:color}), but its location in the
colour-magnitude diagram and its SED are consistent with low surface
gravity, mixed H/He atmosphere WDs.

\begin{figure*}
\centering
\includegraphics[scale=0.6,angle=-90,bb=62 16 576 784]{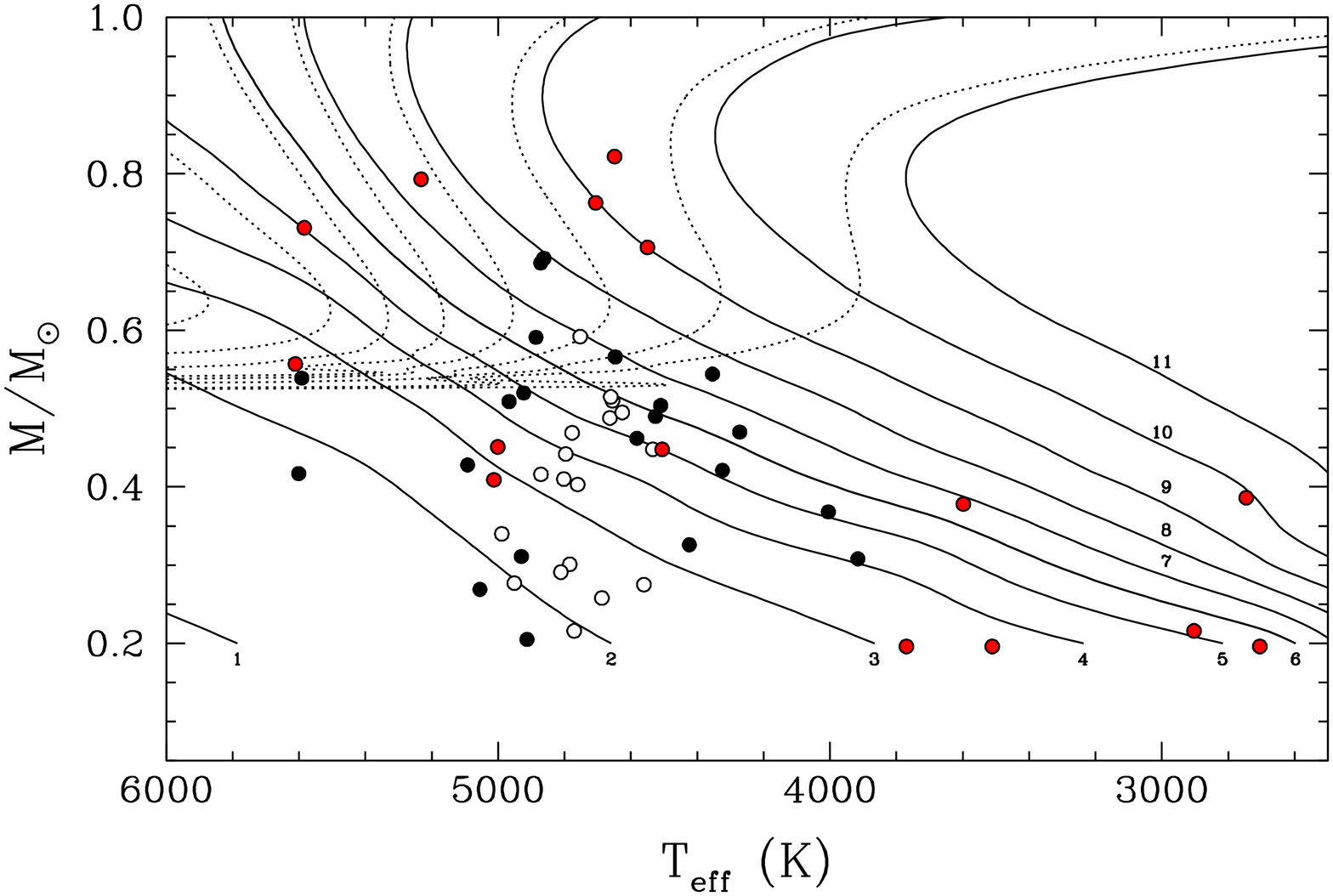}
\caption{Masses of all stars in our sample as a function of
  effective temperature. The black dots correspond to the WDs with
  H-rich atmospheres while the white dots represent the He-rich
  WDs. The red dots represent the 15 cool and ultracool WDs with mixed
  atmospheres. Also shown are theoretical isochrones with their ages
  labelled in Gyr; solid lines correspond to WD cooling ages
  only, while the dotted lines also include the main-sequence
  lifetime.
\label{fg:iso}}
\end{figure*}

Fig.~\ref{fg:J1102} shows our model fits to the J1102 SED using pure
H atmosphere models. We find \Te\ =~3921~K and \logg\ =~8.16 for this
ultracool WD. These parameters are consistent with the analysis based
on the \citet{kowalski06} models within the errors
\citep{kilic12}. Both \citet{kowalski06} and \citet{bergeron95} models
underpredict the flux in the $i$ and $z$ bands for this star. Addition
of He provides a significantly better fit to the SED. The best-fitting
mixed atmosphere model has trace amounts of helium with \Te\ =~3327~K,
\logg\ =~7.64, and $\log{({\rm He/H})}=-3.51$. Based on this, all
three of the previously analysed ultracool WDs with parallax
measurements would be best explained by mixed H/He atmosphere models.

The six ultracool WDs in our sample (see Fig.~\ref{fg:UC}) are all
similar to LHS~3250. They are overluminous compared to pure H
atmosphere WDs, and their SEDs and locations in colour-magnitude
diagrams are matched fairly well by mixed H/He atmosphere WD
models. The models have problems matching the peaks of the energy
distributions and they predict absorption features at 0.8 and 1.1
$\mu$m that are not observed, but the overall fits are quite
reasonable. Based on these, all nine (including the six in our sample)
ultracool WDs with parallax measurements are best explained with
H-rich (mixed) atmospheres. In addition, seven of the nine are
low-mass objects with He-cores. These low-mass objects are about twice
as large (see Table 4), and therefore four times as bright, as typical
0.6~\msun\ WDs. Hence, their overabundance in the SDSS sample is not
surprising.

\subsection{Common Proper Motion Pairs}

There are three common proper motion pairs in our sample. These
include two WD + WD pairs (J0747+2438 and J2239+0018) and one WD + K
dwarf (J0045+1420). The latter was reported as a common proper motion
binary by \citet{luyten87} and \citet{lepine07}. LSPM J0045+1421
(BD+13 99) is a G8V star 62.5 arcsec away from LSPM J0045+1420.  We
confirm that both the WD (J0045+1420) and the G8 dwarf are at the same
distance, making it a Sirius-like binary.  Such binary systems can be
used to constrain the initial--final mass relation.  However,
J0045+1420 is a 0.43~$\pm$~0.07~\msun\ low mass WD with a cooling age
of 2.7 Gyr.  Due to its low-mass, J0045+1420 may itself be an
unresolved binary, and it is impossible to constrain its total age or
the mass of its progenitor star.

The remaining two WD + WD systems are also very useful as they provide
a test of our cooling age estimates. J0747+2438 contains a
$2.5^{+0.5}_{-0.3}$~Gyr old, 0.54~$\pm$~0.05~\msun\ DA WD and a
$4.5^{+1.1}_{-0.9}$~Gyr old, 0.47~$\pm$~0.05~\msun\ DC WD that is
best-fitted by a pure He atmosphere model. The separation between the two
WDs is $\approx$~2000~au; it is safe to assume that both stars evolved
independently. The cooling ages of the two stars differ by $\approx
2\sigma$, and this difference could be due to a difference in the mass
of the progenitor main-sequence stars, though the lower mass star
(J0747+2438N) is also the older star in this system.  The progenitor
stars of these relatively low-mass WDs were likely Sun-like stars that
lived for 10 Gyr, and J0747+2438 is probably a very old binary system
in the Galactic thick disc (it has a tangential velocity of only $38
\pm 2$~\kms).

J2239+0018 consists of a \Te\ =~4420~$\pm$~90~K cool WD with a
\Te\ =~3510~$\pm$~220~K ultracool WD companion. The two WDs are
separated by only 1.85 arcsec, which corresponds to a physical
separation of 155~au. Unfortunately, the relatively large error in
our parallax measurement translates into a large error in mass and
cooling age estimates for this binary. J2239+0018B has a cooling age
of $6.3^{+7.2}_{-3.4}$~Gyr, whereas the ultracool WD J2239+0018A has a
cooling age of $5.2^{+3.7}_{-1.5}$~Gyr. Given the large errors, these
two estimates are consistent within the errors. Further insight into
understanding similar binary systems will require more accurate distance
measurements than those currently available.

\subsection{Disc versus halo}

Fig.~\ref{fg:iso} shows the mass versus temperature distribution of our
parallax sample along with the theoretical isochrones for WDs with CO
core compositions and thick envelopes, i.e., $q({\rm He}) = 10^{-2}$
and $q({\rm H}) = 10^{-4}$. \citet{bergeron01} explain the observed
trend in the isochrones. The general trend is that low-mass WDs evolve
(cool) faster than their counterparts, except that the onset of
crystallization in the most massive WDs shortens the cooling times
considerably. This leads to the parabola shaped isochrones. We also
show the corresponding isochrones for the total main-sequence + WD
cooling ages for $\tau \geqslant 2$~Gyr. We simply assume $t_{\rm
  MS}$~=~10$(M_{\rm MS}/M_{\odot})^{-2.5}$~Gyr and $M_{\rm
  MS}/M_{\odot}$~=~8$\ln[(M_{\rm WD}/M_{\odot})/0.4]$
\citep{wood92,leggett98}.

\begin{figure}
\centering
\includegraphics[scale=0.475,bb=45 177 592 729]{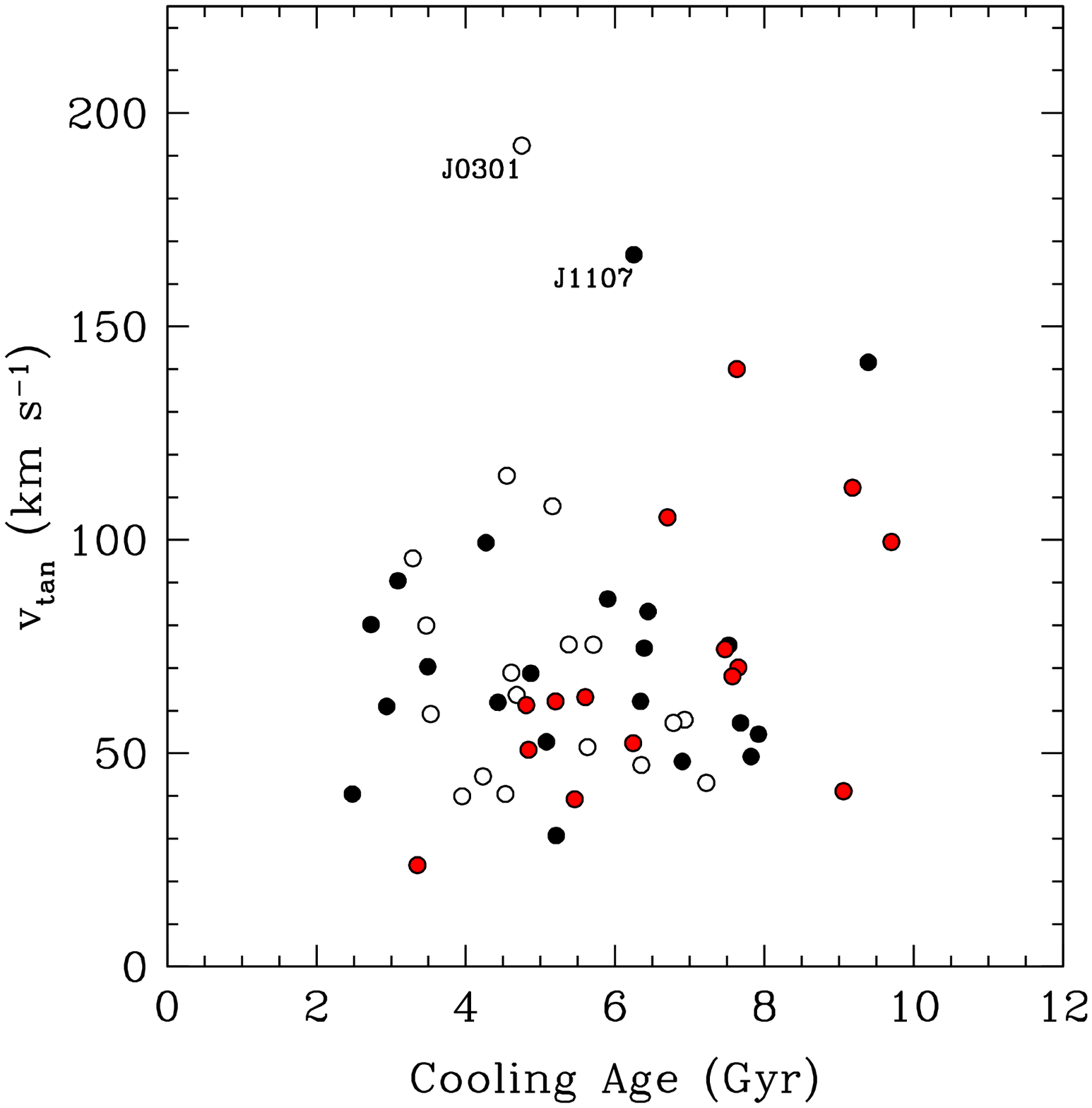}
\caption{The tangential velocity, $v_{\rm tan}$, plotted as a function
  of the WD cooling age. The black dots correspond to the WDs with
  H-rich atmospheres while the white dots represent the He-rich
  WDs. The red dots represent the 15 cool and ultracool WDs with mixed
  atmospheres.
\label{fg:age}}
\end{figure}

The oldest globular clusters in the halo are currently producing
0.53~\msun\ WDs \citep{hansen07}.  There are several WDs in our sample
with $M\approx$~0.53~\msun. J1436+4332 is an excellent example of a
potentially very old star. This DC WD is a \Te\ =~4630~$\pm$~30~K,
\logg\ =~7.88, $M$~=~0.50~$\pm$~0.03~\msun\ He-atmosphere WD with a
WD cooling age of $5.4 \pm 0.6$ Gyr. Its progenitor main-sequence star
was most likely a Sun-like star, with a main-sequence lifetime of 10
Gyr. Hence, the total main-sequence + WD age of J1436+4332 could be as
much as the age of the Universe.

Fig.~\ref{fg:iso} demonstrates that WDs with $M <$~0.5~\msun\ cannot
have CO cores and also form through single star evolution within the
lifetime of the Universe. Hence, they must be either He-core WDs or
unresolved double degenerate systems. In fact, the majority of the
targets in our parallax programme, including all of the ones with
\Te\ $\leq$~4300~K, seem to be low-mass WDs with $M <$~0.5~\msun. The
isochrones shown in this figure are not appropriate for these low-mass
WDs. Based on the \citet{althaus01} models, they have cooling ages
ranging from 2.9 to 9.7 Gyr. Since the prior evolution of these
systems, including the masses of their main-sequence progenitors, is
unknown, their total ages cannot be estimated. However, some of the
low-mass ultracool WDs are clearly very old.

Tables~\ref{tab:astro} and~\ref{tab:tg} and Fig.~\ref{fg:age}
present tangential velocities and cooling ages for our parallax
sample.  The cooling ages range from about 2 to 10~Gyr, whereas the
tangential velocities of all but two of the targets are less than
150~\kms. The only targets that display halo kinematics are
J0301$-$0044 and J1107+4855 (hereafter J0301 and J1107,
respectively). J0301 and J1107 have tangential velocities of
167--192~\kms, \Te\ $\approx$~4500~K, $M$~=~0.45--0.50~\msun and WD
cooling ages of 5--6~Gyr. If both stars are single CO core WDs, their
progenitors would be Sun-like stars with main-sequence lifetimes of
10~Gyr. Hence, their total main-sequence + WD cooling ages would be
$\sim$14~Gyr, which is consistent with a halo origin.

\begin{figure}
\centering
\includegraphics[scale=0.425,bb=30 167 592 679]{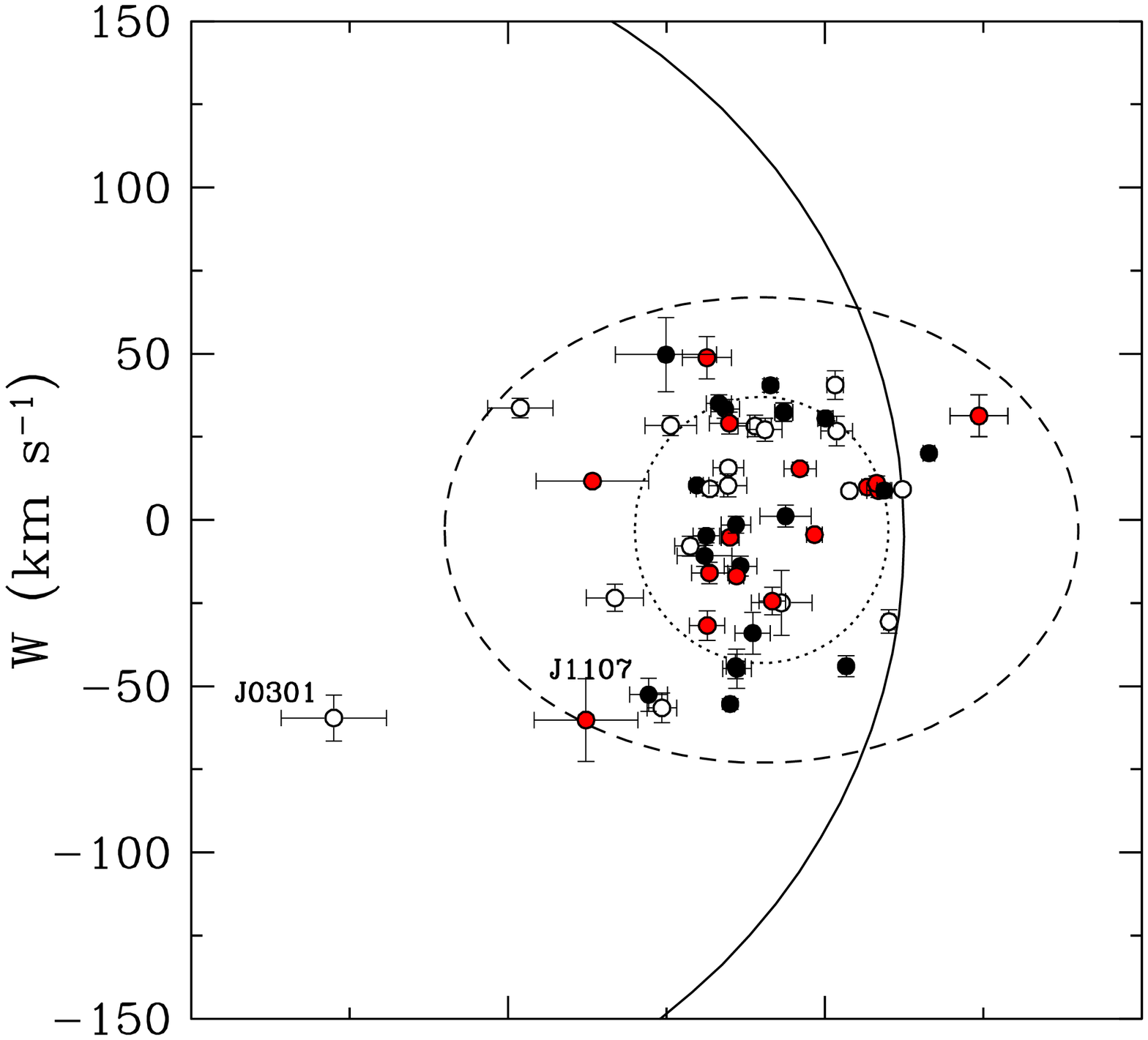}
\includegraphics[scale=0.425,bb=30 117 592 679]{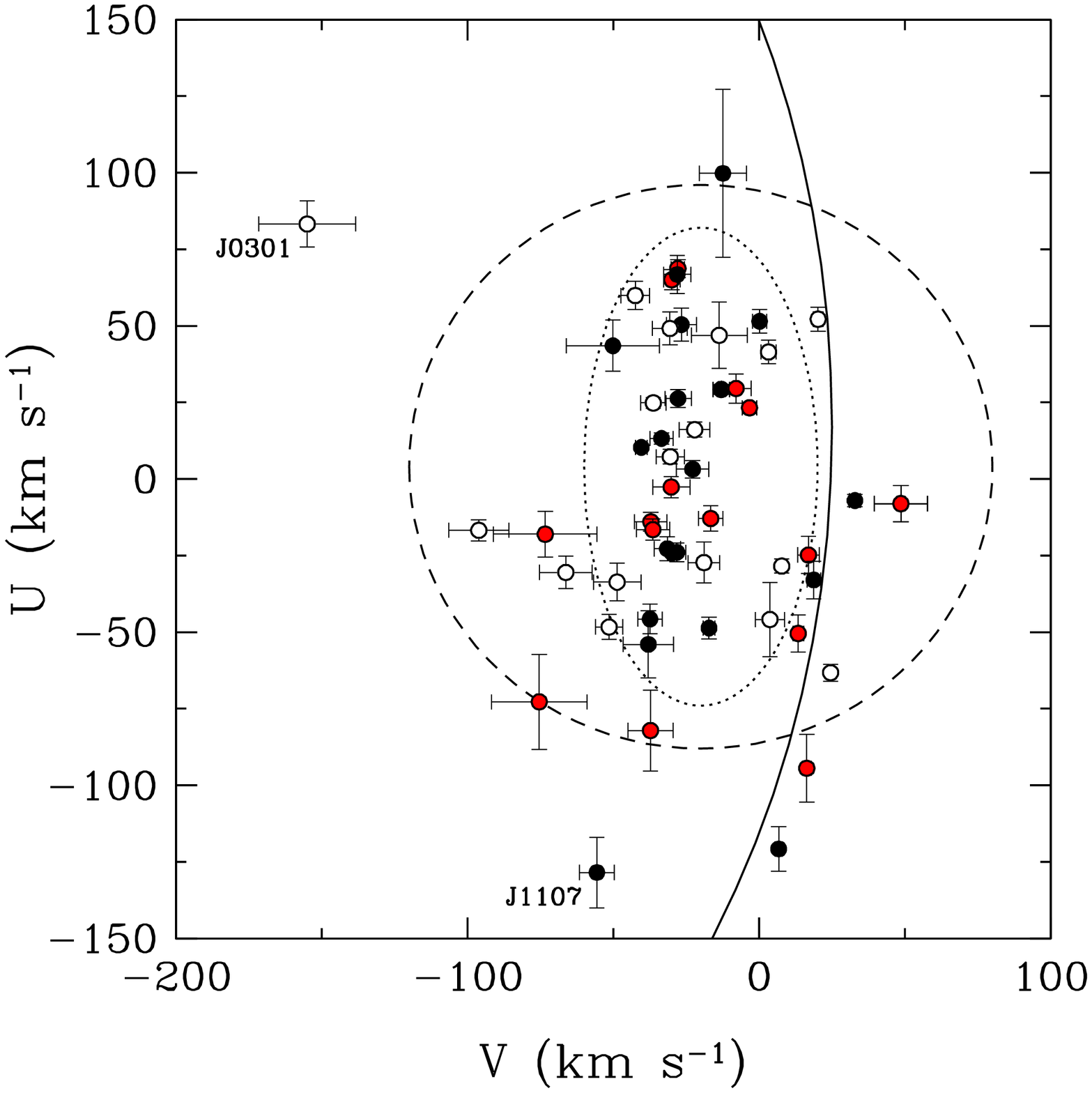}
\caption{\emph{W} versus \emph{V} (top) and \emph{U} versus \emph{V}
  (bottom) velocity distributions for the 54 WDs in our sample. The
  ellipsoids denote the 2$\sigma$ contours for the Galactic thin-disc
  (dotted), thick-disc (dashed) and stellar halo populations (solid).
  The black dots correspond to the WDs with H-rich atmospheres while
  the white dots represent the He-rich WDs. The red dots represent the
  15 cool and ultracool WDs with mixed atmospheres.
\label{fg:UVW}}
\end{figure}

In Table~\ref{tab:astro}, the last three columns list the $(U,V,W)$
components of the space velocities for each WD. These space velocities
have been computed by combining the observed parallaxes and proper
motions for each WD using the prescription of \citet{johnson87}. Since
we do not have any radial velocity measurements for our WDs, we assume
a radial velocity $v_{\rm rad}$~=~0~\kms\ in the calculation of
$U$. In Fig.~\ref{fg:UVW}, we plot the resulting $W$ versus $V$ (top),
and $U$ versus $V$ (bottom) velocity distributions. We also include the
2$\sigma$ contours for the Galactic thin disc (dotted), thick disc
(dashed) and stellar halo populations (solid) \citep{chiba00}. It is
clear from Figure~\ref{fg:UVW} that both J0301 and J1107 are most
likely halo WDs. The distribution of the remaining sample, including
the IR-faint (the DC WDs with mixed H/He atmospheres) and ultracool
WDs, is consistent with disc membership. The ratio of thick to thin
disc WDs is 18/34. If we assume that J0301 and J1107 are indeed bona
fide members of the halo, the observed velocity distribution suggests
63 per cent/33 per cent/4 per cent proportions for the contribution of
thin disc, thick disc and halo WDs for our sample.

We note that this is the first time a large number of ultracool WDs
have distance and tangential velocities available, and in contrast to
the expectations, they all seem to be members of the disc. These WDs
provide an independent constraint on the thick disc population; the
Galactic thick disc is at least 10~Gyr old.

\section{Conclusions}

We present parallax observations of 54 cool and ultracool WDs.  Our
sample includes one new WD within the local 25~pc sample and five
stars within 30 pc. All but two of them have tangential velocities
smaller than 150~\kms. J0301$-$0044 and J1107+4855 are the only
objects in our sample with kinematics and ages that are consistent
with halo WDs. The rest of the objects, including the ultracool WDs,
are members of the Galactic thick disc. The oldest WDs in this sample
have WD cooling ages of 10 Gyr, providing a firm lower limit to the
age of the thick disc. Many of our targets are low-mass WDs. These are
either single He-core WDs or unresolved double degenerates. It appears
that we have detected the brighter population of cool and ultracool
WDs in the solar neighbourhood, and the fainter, normal CO core
ultracool WDs remain to be discovered in large numbers. Future and
upcoming astrometric surveys such as the LSST will find those
fainter and more massive ultracool WDs.

\section*{Acknowledegments}

We would like to thank the anonymous referee for a careful reading 
of our manuscript and several constructive comments that helped improve 
this paper. We thank Hugh C. Harris for useful discussions and suggestions. AG
acknowledges support provided by NASA through grant number
HST-GO-13319.01 from the Space Telescope Science Institute, which is
operated by AURA, Inc., under NASA contract NAS 5-26555. MK gratefully
acknowledges support from the NSF and NASA under grants AST-1312678
and NNX14AF65G, respectively. JRT acknowledges support from the
NSF under grants AST-0708810 and AST-1008217. This work was supported
in part by the NSERC Canada and by the Fund FRQ-NT (Qu\'ebec). This
research makes use of the SAO/NASA Astrophysics Data System
Bibliographic Service. This project makes use of data products from
the SDSS, which has been funded by the Alfred
P. Sloan Foundation, the Participating Institutions, the National
Science Foundation, and the US Department of Energy Office of
Science.

\bibliographystyle{mn2e}
\bibliography{biblio}

\begin{thebibliography}{56}
\expandafter\ifx\csname natexlab\endcsname\relax\def\natexlab#1{#1}\fi

\bibitem[{{Ahn} {et~al}\mbox{.}(2014){Ahn}, {Alexandroff}, {Allende Prieto},
  {Anders}, {Anderson}, {Anderton}, {Andrews}, {Aubourg}, {Bailey}, {Bastien},
  \& et~al.}]{ahn14}
{Ahn} C.~P. {et~al.}, 2014, ApJS, 211, 17

\bibitem[{{Althaus}, {Serenelli} \& {Benvenuto}(2001){Althaus}, {Serenelli}, \&
  {Benvenuto}}]{althaus01}
{Althaus} L.~G., {Serenelli} A.~M., {Benvenuto} O.~G., 2001, MNRAS, 323, 471

\bibitem[{{Bergeron} \& {Leggett}(2002)}]{bergeron02}
{Bergeron} P., {Leggett} S.~K., 2002, ApJ, 580, 1070

\bibitem[{{Bergeron}, {Leggett} \& {Ruiz}(2001){Bergeron}, {Leggett}, \&
  {Ruiz}}]{bergeron01}
{Bergeron} P., {Leggett} S.~K., {Ruiz} M.~T., 2001, ApJS, 133, 413

\bibitem[{{Bergeron}, {Ruiz} \& {Leggett}(1997){Bergeron}, {Ruiz}, \&
  {Leggett}}]{brl97}
{Bergeron} P., {Ruiz} M.~T., {Leggett} S.~K., 1997, ApJS, 108, 339

\bibitem[{{Bergeron}, {Saumon} \& {Wesemael}(1995){Bergeron}, {Saumon}, \&
  {Wesemael}}]{bergeron95}
{Bergeron} P., {Saumon} D., {Wesemael} F., 1995, ApJ, 443, 764

\bibitem[{{Bergeron} {et~al}\mbox{.}(2011){Bergeron}, {Wesemael}, {Dufour},
  {Beauchamp}, {Hunter}, {Saffer}, {Gianninas}, {Ruiz}, {Limoges}, {Dufour},
  {Fontaine}, \& {Liebert}}]{bergeron11}
{Bergeron} P. {et~al.}, 2011, ApJ, 737, 28

\bibitem[{{Bohlin} \& {Gilliland}(2004)}]{bohlin04}
{Bohlin} R.~C., {Gilliland} R.~L., 2004, AJ, 127, 3508

\bibitem[{{Brown} {et~al}\mbox{.}(2011){Brown}, {Kilic}, {Brown}, \&
  {Kenyon}}]{brown11}
{Brown} J.~M., {Kilic} M., {Brown} W.~R., {Kenyon} S.~J., 2011, ApJ, 730, 67

\bibitem[{{Catal{\'a}n} {et~al}\mbox{.}(2012){Catal{\'a}n}, {Tremblay},
  {Pinfield}, {Smith}, {Zhang}, {Napiwotzki}, {Marocco}, {Day-Jones}, {Gomes},
  {Forde}, {Lucas}, \& {Jones}}]{catalan12}
{Catal{\'a}n} S. {et~al.}, 2012, A\&A, 546, L3

\bibitem[{{Chiba} \& {Beers}(2000)}]{chiba00}
{Chiba} M., {Beers} T.~C., 2000, AJ, 119, 2843

\bibitem[{{Fontaine}, {Brassard} \& {Bergeron}(2001){Fontaine}, {Brassard}, \&
  {Bergeron}}]{fontaine01}
{Fontaine} G., {Brassard} P., {Bergeron} P., 2001, PASP, 113, 409

\bibitem[{{Gates} {et~al}\mbox{.}(2004){Gates}, {Gyuk}, {Harris}, {Subbarao},
  {Anderson}, {Kleinman}, {Liebert}, {Brewington}, {Brinkmann}, {Harvanek},
  {Krzesinski}, {Lamb}, {Long}, {Neilsen}, {Newman}, {Nitta}, \&
  {Snedden}}]{gates04}
{Gates} E. {et~al.}, 2004, ApJ, 612, L129

\bibitem[{{Giammichele}, {Bergeron} \& {Dufour}(2012){Giammichele}, {Bergeron},
  \& {Dufour}}]{giam12}
{Giammichele} N., {Bergeron} P., {Dufour} P., 2012, ApJS, 199, 29

\bibitem[{{Hall} {et~al}\mbox{.}(2008){Hall}, {Kowalski}, {Harris}, {Awal},
  {Leggett}, {Kilic}, {Anderson}, \& {Gates}}]{hall08}
{Hall} P.~B., {Kowalski} P.~M., {Harris} H.~C., {Awal} A., {Leggett} S.~K.,
  {Kilic} M., {Anderson} S.~F., {Gates} E., 2008, AJ, 136, 76

\bibitem[{{Hambly} {et~al}\mbox{.}(1999){Hambly}, {Smartt}, {Hodgkin},
  {Jameson}, {Kemp}, {Rolleston}, \& {Steele}}]{hambly99}
{Hambly} N.~C., {Smartt} S.~J., {Hodgkin} S.~T., {Jameson} R.~F., {Kemp} S.~N.,
  {Rolleston} W.~R.~J., {Steele} I.~A., 1999, MNRAS, 309, L33

\bibitem[{{Hansen}(1999)}]{hansen99}
{Hansen} B.~M.~S., 1999, ApJ, 520, 680

\bibitem[{{Hansen} {et~al}\mbox{.}(2007){Hansen}, {Anderson}, {Brewer},
  {Dotter}, {Fahlman}, {Hurley}, {Kalirai}, {King}, {Reitzel}, {Richer},
  {Rich}, {Shara}, \& {Stetson}}]{hansen07}
{Hansen} B.~M.~S. {et~al.}, 2007, ApJ, 671, 380

\bibitem[{{Hansen} {et~al}\mbox{.}(2004){Hansen}, {Richer}, {Fahlman},
  {Stetson}, {Brewer}, {Currie}, {Gibson}, {Ibata}, {Rich}, \&
  {Shara}}]{hansen04}
{Hansen} B.~M.~S. {et~al.}, 2004, ApJS, 155, 551

\bibitem[{{Harris} {et~al}\mbox{.}(1999){Harris}, {Dahn}, {Vrba}, {Henden},
  {Liebert}, {Schmidt}, \& {Reid}}]{harris99}
{Harris} H.~C., {Dahn} C.~C., {Vrba} F.~J., {Henden} A.~A., {Liebert} J.,
  {Schmidt} G.~D., {Reid} I.~N., 1999, ApJ, 524, 1000

\bibitem[{{Harris} {et~al}\mbox{.}(2008){Harris}, {Gates}, {Gyuk}, {Subbarao},
  {Anderson}, {Hall}, {Munn}, {Liebert}, {Knapp}, {Bizyaev}, {Malanushenko},
  {Malanushenko}, {Pan}, {Schneider}, \& {Smith}}]{harris08}
{Harris} H.~C. {et~al.}, 2008, ApJ, 679, 697

\bibitem[{{Harris} {et~al}\mbox{.}(2006){Harris}, {Munn}, {Kilic}, {Liebert},
  {Williams}, {von Hippel}, {Levine}, {Monet}, {Eisenstein}, {Kleinman},
  {Metcalfe}, {Nitta}, {Winget}, {Brinkmann}, {Fukugita}, {Knapp}, {Lupton},
  {Smith}, \& {Schneider}}]{harris06}
{Harris} H.~C. {et~al.}, 2006, AJ, 131, 571

\bibitem[{{Holberg} \& {Bergeron}(2006)}]{hb06}
{Holberg} J.~B., {Bergeron} P., 2006, AJ, 132, 1221

\bibitem[{{Holberg} {et~al}\mbox{.}(2008){Holberg}, {Sion}, {Oswalt}, {McCook},
  {Foran}, \& {Subasavage}}]{holberg08}
{Holberg} J.~B., {Sion} E.~M., {Oswalt} T., {McCook} G.~P., {Foran} S.,
  {Subasavage} J.~P., 2008, AJ, 135, 1225

\bibitem[{{Hummer} \& {Mihalas}(1988)}]{hm88}
{Hummer} D.~G., {Mihalas} D., 1988, ApJ, 331, 794

\bibitem[{{Iben} \& {Tutukov}(1984)}]{iben84}
{Iben}, Jr. I., {Tutukov} A.~V., 1984, ApJ, 282, 615

\bibitem[{{Ivezic} {et~al}\mbox{.}(2008){Ivezic}, {Tyson}, {Abel}, {Acosta},
  {Allsman}, {AlSayyad}, {Anderson}, {Andrew}, {Angel}, {Angeli}, {Ansari},
  {Antilogus}, {Arndt}, {Astier}, {Aubourg}, {Axelrod}, {Bard}, {Barr},
  {Barrau}, {Bartlett}, {Bauman}, {Beaumont}, {Becker}, {Becla}, {Beldica},
  {Bellavia}, {Blanc}, {Blandford}, {Bloom}, {Bogart}, {Borne}, {Bosch},
  {Boutigny}, {Brandt}, {Brown}, {Bullock}, {Burchat}, {Burke}, {Cagnoli},
  {Calabrese}, {Chandrasekharan}, {Chesley}, {Cheu}, {Chiang}, {Claver},
  {Connolly}, {Cook}, {Cooray}, {Covey}, {Cribbs}, {Cui}, {Cutri}, {Daubard},
  {Daues}, {Delgado}, {Digel}, {Doherty}, {Dubois}, {Dubois-Felsmann},
  {Durech}, {Eracleous}, {Ferguson}, {Frank}, {Freemon}, {Gangler}, {Gawiser},
  {Geary}, {Gee}, {Geha}, {Gibson}, {Gilmore}, {Glanzman}, {Goodenow},
  {Gressler}, {Gris}, {Guyonnet}, {Hascall}, {Haupt}, {Hernandez}, {Hogan},
  {Huang}, {Huffer}, {Innes}, {Jacoby}, {Jain}, {Jee}, {Jernigan},
  {Jevremovic}, {Johns}, {Jones}, {Juramy-Gilles}, {Juric}, {Kahn}, {Kalirai},
  {Kallivayalil}, {Kalmbach}, {Kantor}, {Kasliwal}, {Kessler}, {Kirkby},
  {Knox}, {Kotov}, {Krabbendam}, {Krughoff}, {Kubanek}, {Kuczewski},
  {Kulkarni}, {Lambert}, {Le Guillou}, {Levine}, {Liang}, {Lim}, {Lintott},
  {Lupton}, {Mahabal}, {Marshall}, {Marshall}, {May}, {McKercher}, {Migliore},
  {Miller}, {Mills}, {Monet}, {Moniez}, {Neill}, {Nief}, {Nomerotski},
  {Nordby}, {O'Connor}, {Oliver}, {Olivier}, {Olsen}, {Ortiz}, {Owen}, {Pain},
  {Peterson}, {Petry}, {Pierfederici}, {Pietrowicz}, {Pike}, {Pinto}, {Plante},
  {Plate}, {Price}, {Prouza}, {Radeka}, {Rajagopal}, {Rasmussen}, {Regnault},
  {Ridgway}, {Ritz}, {Rosing}, {Roucelle}, {Rumore}, {Russo}, {Saha},
  {Sassolas}, {Schalk}, {Schindler}, {Schneider}, {Schumacher}, {Sebag},
  {Sembroski}, {Seppala}, {Shipsey}, {Silvestri}, {Smith}, {Smith}, {Strauss},
  {Stubbs}, {Sweeney}, {Szalay}, {Takacs}, {Thaler}, {Van Berg}, {Vanden Berk},
  {Vetter}, {Virieux}, {Xin}, {Walkowicz}, {Walter}, {Wang}, {Warner},
  {Willman}, {Wittman}, {Wolff}, {Wood-Vasey}, {Yoachim}, {Zhan}, \& {for the
  LSST Collaboration}}]{ivezic08}
{Ivezic} Z. {et~al.}, 2008, preprint (arXiv:0805.2366)

\bibitem[{{Johnson} \& {Soderblom}(1987)}]{johnson87}
{Johnson} D.~R.~H., {Soderblom} D.~R., 1987, AJ, 93, 864

\bibitem[{{Kilic} {et~al}\mbox{.}(2010){Kilic}, {Leggett}, {Tremblay}, {von
  Hippel}, {Bergeron}, {Harris}, {Munn}, {Williams}, {Gates}, \&
  {Farihi}}]{kilic10a}
{Kilic} M. {et~al.}, 2010, ApJS, 190, 77

\bibitem[{{Kilic} {et~al}\mbox{.}(2006){Kilic}, {Munn}, {Harris}, {Liebert},
  {von Hippel}, {Williams}, {Metcalfe}, {Winget}, \& {Levine}}]{kilic06a}
{Kilic} M. {et~al.}, 2006, AJ, 131, 582

\bibitem[{{Kilic} {et~al}\mbox{.}(2012){Kilic}, {Thorstensen}, {Kowalski}, \&
  {Andrews}}]{kilic12}
{Kilic} M., {Thorstensen} J.~R., {Kowalski} P.~M., {Andrews} J., 2012, MNRAS,
  423, L132

\bibitem[{{Kowalski} \& {Saumon}(2006)}]{kowalski06}
{Kowalski} P.~M., {Saumon} D., 2006, ApJ, 651, L137

\bibitem[{{Lawrence} {et~al}\mbox{.}(2007){Lawrence}, {Warren}, {Almaini},
  {Edge}, {Hambly}, {Jameson}, {Lucas}, {Casali}, {Adamson}, {Dye}, {Emerson},
  {Foucaud}, {Hewett}, {Hirst}, {Hodgkin}, {Irwin}, {Lodieu}, {McMahon},
  {Simpson}, {Smail}, {Mortlock}, \& {Folger}}]{lawrence07}
{Lawrence} A. {et~al.}, 2007, MNRAS, 379, 1599

\bibitem[{{Leggett} {et~al}\mbox{.}(2011){Leggett}, {Lodieu}, {Tremblay},
  {Bergeron}, \& {Nitta}}]{leggett11}
{Leggett} S.~K., {Lodieu} N., {Tremblay} P.-E., {Bergeron} P., {Nitta} A.,
  2011, ApJ, 735, 62

\bibitem[{{Leggett}, {Ruiz} \& {Bergeron}(1998){Leggett}, {Ruiz}, \&
  {Bergeron}}]{leggett98}
{Leggett} S.~K., {Ruiz} M.~T., {Bergeron} P., 1998, ApJ, 497, 294

\bibitem[{{L{\'e}pine} \& {Bongiorno}(2007)}]{lepine07}
{L{\'e}pine} S., {Bongiorno} B., 2007, AJ, 133, 889

\bibitem[{{L{\'e}pine} \& {Shara}(2005)}]{lepine05}
{L{\'e}pine} S., {Shara} M.~M., 2005, AJ, 129, 1483

\bibitem[{{Liebert}, {Dahn} \& {Monet}(1988){Liebert}, {Dahn}, \&
  {Monet}}]{liebert88}
{Liebert} J., {Dahn} C.~C., {Monet} D.~G., 1988, ApJ, 332, 891

\bibitem[{{Luyten}(1987)}]{luyten87}
{Luyten} W.~J., 1987, Proper Motion Survey with Schmidt Telescopes, 71, 1

\bibitem[{{Marsh}, {Dhillon} \& {Duck}(1995){Marsh}, {Dhillon}, \&
  {Duck}}]{marsh95}
{Marsh} T.~R., {Dhillon} V.~S., {Duck} S.~R., 1995, MNRAS, 275, 828

\bibitem[{{Mestel}(1952)}]{mestel52}
{Mestel} L., 1952, MNRAS, 112, 583

\bibitem[{{Munn} {et~al}\mbox{.}(2004){Munn}, {Monet}, {Levine}, {Canzian},
  {Pier}, {Harris}, {Lupton}, {Ivezi{\'c}}, {Hindsley}, {Hennessy},
  {Schneider}, \& {Brinkmann}}]{munn04}
{Munn} J.~A. {et~al.}, 2004, AJ, 127, 3034

\bibitem[{{Oppenheimer} {et~al}\mbox{.}(2001){Oppenheimer}, {Saumon},
  {Hodgkin}, {Jameson}, {Hambly}, {Chabrier}, {Filippenko}, {Coil}, \&
  {Brown}}]{oppen01}
{Oppenheimer} B.~R. {et~al.}, 2001, ApJ, 550, 448

\bibitem[{{Perryman} {et~al}\mbox{.}(2001){Perryman}, {de Boer}, {Gilmore},
  {H{\o}g}, {Lattanzi}, {Lindegren}, {Luri}, {Mignard}, {Pace}, \& {de
  Zeeuw}}]{perryman01}
{Perryman} M.~A.~C. {et~al.}, 2001, A\&A, 369, 339

\bibitem[{{Press} {et~al}\mbox{.}(1986){Press}, {Flannery}, {Teukolsky}, \&
  {Vetterling}}]{press86}
{Press} W.~H., {Flannery} B.~P., {Teukolsky} S.~A., {Vetterling} W.~T., 1986,
  {Numerical Recipes}

\bibitem[{{Rowell}, {Kilic} \& {Hambly}(2008){Rowell}, {Kilic}, \&
  {Hambly}}]{rowell08}
{Rowell} N.~R., {Kilic} M., {Hambly} N.~C., 2008, MNRAS, 385, L23

\bibitem[{{Sion} {et~al}\mbox{.}(2014){Sion}, {Holberg}, {Oswalt}, {McCook},
  {Wasatonic}, \& {Myszka}}]{sion14}
{Sion} E.~M., {Holberg} J.~B., {Oswalt} T.~D., {McCook} G.~P., {Wasatonic} R.,
  {Myszka} J., 2014, AJ, 147, 129

\bibitem[{{Skrutskie} {et~al}\mbox{.}(2006){Skrutskie}, {Cutri}, {Stiening},
  {Weinberg}, {Schneider}, {Carpenter}, {Beichman}, {Capps}, {Chester},
  {Elias}, {Huchra}, {Liebert}, {Lonsdale}, {Monet}, {Price}, {Seitzer},
  {Jarrett}, {Kirkpatrick}, {Gizis}, {Howard}, {Evans}, {Fowler}, {Fullmer},
  {Hurt}, {Light}, {Kopan}, {Marsh}, {McCallon}, {Tam}, {Van Dyk}, \&
  {Wheelock}}]{skrut06}
{Skrutskie} M.~F. {et~al.}, 2006, AJ, 131, 1163

\bibitem[{{Thorstensen}(2003)}]{thor03}
{Thorstensen} J.~R., 2003, AJ, 126, 3017

\bibitem[{{Thorstensen}, {L{\'e}pine} \& {Shara}(2008){Thorstensen},
  {L{\'e}pine}, \& {Shara}}]{thor08}
{Thorstensen} J.~R., {L{\'e}pine} S., {Shara} M., 2008, AJ, 136, 2107

\bibitem[{{Tokunaga}, {Simons} \& {Vacca}(2002){Tokunaga}, {Simons}, \&
  {Vacca}}]{tokunaga02}
{Tokunaga} A.~T., {Simons} D.~A., {Vacca} W.~D., 2002, PASP, 114, 180

\bibitem[{{Tremblay} \& {Bergeron}(2009)}]{tb09}
{Tremblay} P.-E., {Bergeron} P., 2009, ApJ, 696, 1755

\bibitem[{{Tremblay} {et~al}\mbox{.}(2014){Tremblay}, {Leggett}, {Lodieu},
  {Freytag}, {Bergeron}, {Kalirai}, \& {Ludwig}}]{tremblay14}
{Tremblay} P.-E., {Leggett} S.~K., {Lodieu} N., {Freytag} B., {Bergeron} P.,
  {Kalirai} J.~S., {Ludwig} H.-G., 2014, ApJ, 788, 103

\bibitem[{{Vidrih} {et~al}\mbox{.}(2007){Vidrih}, {Bramich}, {Hewett}, {Evans},
  {Gilmore}, {Hodgkin}, {Smith}, {Wyrzykowski}, {Belokurov}, {Fellhauer},
  {Irwin}, {McMahon}, {Zucker}, {Munn}, {Lin}, {Miknaitis}, {Harris}, {Lupton},
  \& {Schneider}}]{vidrih07}
{Vidrih} S. {et~al.}, 2007, MNRAS, 382, 515

\bibitem[{{Winget} {et~al}\mbox{.}(1987){Winget}, {Hansen}, {Liebert}, {van
  Horn}, {Fontaine}, {Nather}, {Kepler}, \& {Lamb}}]{winget87}
{Winget} D.~E., {Hansen} C.~J., {Liebert} J., {van Horn} H.~M., {Fontaine} G.,
  {Nather} R.~E., {Kepler} S.~O., {Lamb} D.~Q., 1987, ApJ, 315, L77

\bibitem[{{Wood}(1992)}]{wood92}
{Wood} M.~A., 1992, ApJ, 386, 539

\end{thebibliography}

\bsp

\end{document}